\documentclass[preprint,aps,prd,10pt,nofootinbib,superscriptaddress,preprintnumbers,titlepage,amssymb,amsmath]{revtex4-2}

\usepackage{graphicx}
\usepackage{amssymb,amsfonts,amsmath}
\usepackage{hyperref}
\usepackage{nicefrac}
\usepackage[utf8]{inputenc}
\usepackage{textcomp}

\graphicspath{{plots/}}
\newcommand\BERN{\affiliation{Albert Einstein Center, Institute for Theoretical Physics, University of Bern, Bern, Switzerland}}
\newcommand\BETHE{\affiliation{Bethe Center for Theoretical Physics, University of Bonn, Bonn, Germany}}
\newcommand\CERN{\affiliation{Theoretical Physics Department, CERN, Esplanade des Particules 1, 1211 Geneva 23, Switzerland}}
\newcommand\CYI{\affiliation{Computation-based Science and Technology Research Center, The Cyprus Institute, Nicosia, Cyprus}}
\newcommand\HISKP{\affiliation{Helmholtz-Institut für Strahlen- und Kernphysik, University of Bonn, Bonn, Germany}}
\newcommand\PRISMAKPH{\affiliation{PRISMA$^+$~Cluster~of~Excellence and Institut~f\"ur~Kernphysik, Johannes~Gutenberg-Universität~Mainz, 55099~Mainz, Germany}}

\begin{document}
\renewcommand{\l}{\left}
\renewcommand{\r}{\right}
\renewcommand{\vec}{\textbf}
\newcommand{\csw}{c_\mathrm{sw}}
\newcommand{\g}[1]{\gamma_{#1}} 
\newcommand{\G}[1]{\gamma^{#1}} 
\newcommand{\DB}[1]{\stackrel{\leftarrow}{D}_{#1}} 
\newcommand{\DF}[1]{\stackrel{\rightarrow}{D}_{#1}} 
\newcommand{\DBF}[1]{\stackrel{\leftrightarrow}{D}_{#1}} 
\newcommand{\trace}[1]{\mathrm{tr}\left[ #1 \right]} 
\newcommand{\tr}{\mathrm{tr}}
\newcommand{\diag}{\mathrm{diag}}  
\newcommand{\bra}[1]{\left< #1 \right|} 
\newcommand{\ket}[1]{\left| #1 \right>} 
\renewcommand{\Re}{\mathrm{Re} \,}
\renewcommand{\Im}{\mathrm{Im} \,}
\newcommand{\chiral}[1]{\mathring{#1}} 
\newcommand{\gev}{\,\mathrm{GeV}}
\newcommand{\kev}{\,\mathrm{keV}}
\newcommand{\mev}{\,\mathrm{MeV}}
\newcommand{\fm}{\,\mathrm{fm}}
\newcommand{\link}[2]{\mathcal{U}_{#2}({#1})}
\newcommand{\Link}[2]{\mathcal{U}({#1};{#2})}
\newcommand{\order}[1]{\mathcal{O}\l({#1}\r)}
\newcommand{\SU}[1]{\mathrm{SU}\l(#1\r)}
\newcommand{\U}[1]{\mathrm{U}\l(#1\r)}
\newcommand{\MSbar}{\overline{\mathrm{MS}}}
\newcommand{\experiment}{\mathrm{exp}}
\newcommand{\phys}{\mathrm{phys}}
\newcommand{\stat}{\mathrm{stat}}
\newcommand{\sys}{\mathrm{sys}}
\newcommand{\syserr}[1]{(#1)_\mathrm{sys}}
\newcommand{\staterr}[1]{(#1)_\mathrm{stat}}
\newcommand{\totalerr}[1]{[#1]_\mathrm{total}}
\newcommand{\chierr}[1]{(#1)_\chi}
\newcommand{\conterr}[1]{(#1)_\mathrm{cont}}
\newcommand{\FSerr}[1]{(#1)_\mathrm{FS}}
\newcommand{\CCFerr}[1]{(#1)_\mathrm{CCF}}
\newcommand{\tins}{t_\mathrm{ins}}
\newcommand{\tsep}{t_\mathrm{sep}}
\newcommand{\MPS}{M_\mathrm{PS}}
\newcommand{\Mpi}{M_\pi}
\newcommand{\MK}{M_\mathrm{K}}
\newcommand{\Meta}{M_\eta}
\newcommand{\Metap}{M_{\eta^\prime}}
\newcommand{\etap}{\eta^\prime}
\newcommand{\fPS}{f_\mathrm{PS}}
\newcommand{\psibar}{\overline{\psi}}
\newcommand{\chibar}{\overline{\chi}}
\newcommand{\Ctwopt}[3]{C_\mathcal{#1}^\mathrm{2pt}(#2, #3)}
\newcommand{\Cthreept}[5]{C_\mathcal{#1}^\mathrm{3pt}(#2,#3,#4,#5)}
\newcommand{\vp}{\vec{p}}
\newcommand{\vpi}{\vec{p}_i}
\newcommand{\vq}{\vec{q}}
\newcommand{\vpf}{\vec{p}_f}
\newcommand{\vxi}{\vec{x}_i}
\newcommand{\vxop}{\vec{x}_{op}}
\newcommand{\vxf}{\vec{x}_f}
\newcommand{\NOTE}[2]{\begin{center}{\color{red}{\fbox{\parbox{0.9\linewidth}{\bf #1: #2}}}}\end{center}\par}

\title{\textbf{\texorpdfstring{$\eta$, $\etap$}{eta,etap} mesons from lattice QCD in fully physical conditions}}

\author{Konstantin~Ottnad}\email{kottnad@uni-mainz.de} \PRISMAKPH
\author{Simone~Bacchio}\CYI
\author{Jacob~Finkenrath}\CERN
\author{Bartosz~Kostrzewa}\HISKP \BETHE
\author{Marcus~Petschlies}\HISKP \BETHE
\author{Ferenc~Pittler}\CYI
\author{Carsten~Urbach}\HISKP \BETHE
\author{Urs~Wenger}\BERN

\date{ \today \\
 \begin{center}
  \includegraphics[draft=false,width=.25\linewidth]{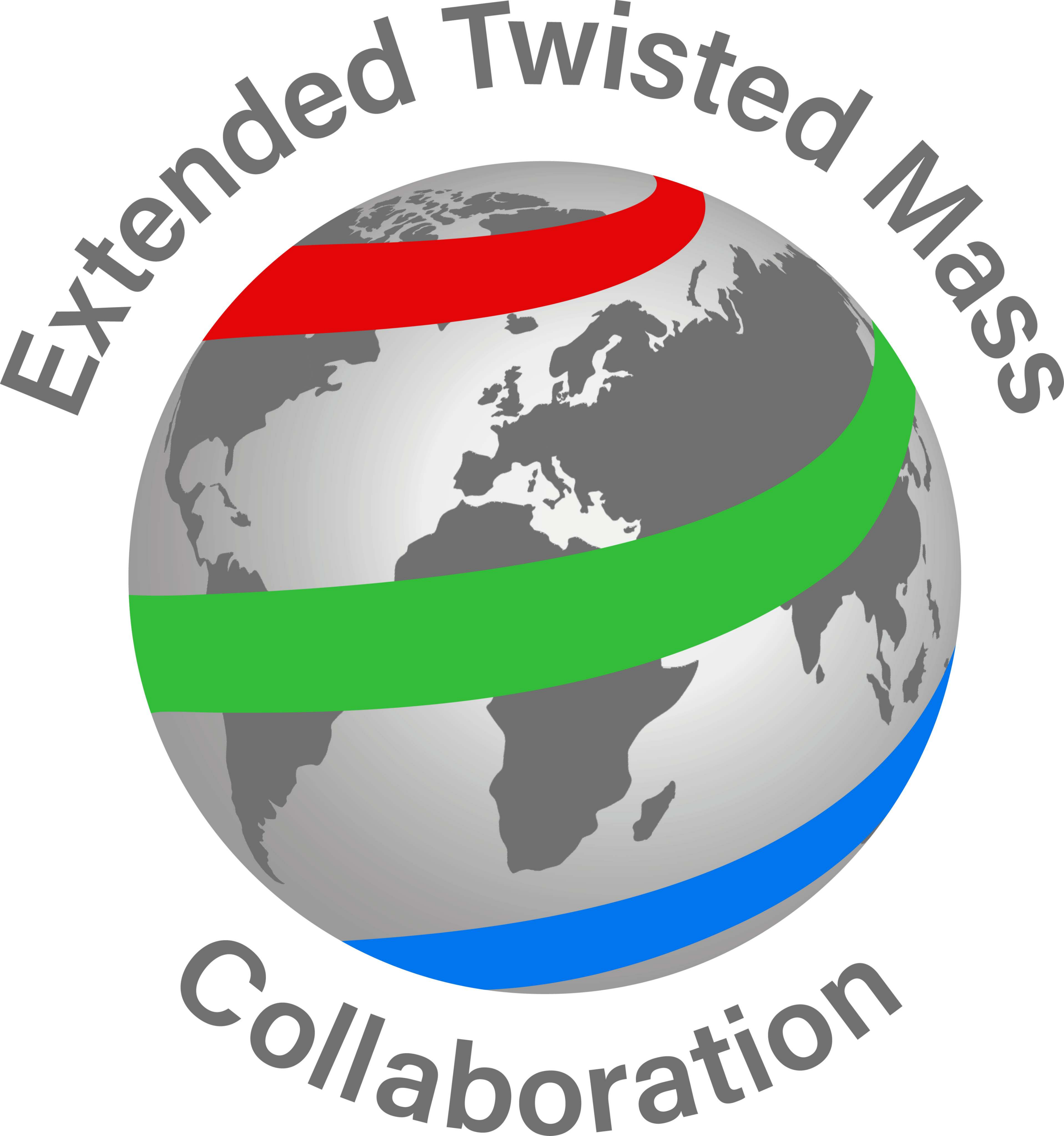}
 \end{center}
}

\preprint{MITP-25-021}
\preprint{CERN-TH-2025-052}

\begin{abstract}
 We determine masses and mixing parameters of the $\eta$ and $\etap$ meson in lattice QCD. The calculations are carried out on a set of 13 ETMC gauge ensembles with $N_f=2+1+1$ (maximally) twisted-mass Clover-improved quarks. These ensemble cover four values of the lattice spacing $a=0.057\fm,...,0.092\fm$ and pion masses from $140\mev$ to $360\mev$, including three ensembles at physical quark masses and six ensembles with $M_\pi<200\mev$. The strange-quark contribution is treated in a mixed-action approach using Osterwalder-Seiler fermions to avoid complications due to flavor mixing in the heavy quark sector and to enable the use of the one-end trick in the computation of strange quark-disconnected diagrams. With the strange-quark mass tuned to its physical value and several ensembles having close-to-physical light-quark mass, uncertainties related to the chiral extrapolations are reduced significantly compared to earlier studies. Physical results are computed with fully controlled systematics from a combined chiral, continuum and infinite-volume extrapolation, and a full error budget is obtained from model averages over of various fit ans\"atze and data cuts. Our results for the masses are given by $\Meta=551(16)\mev$ and $\Metap=972(20)\mev$, respectively, where statistical and systematic errors have been added in quadrature. For the mixing angle and decay-constant parameters the Feldmann-Kroll-Stech scheme is employed to compute them from pseudoscalar matrix elements in the quark-flavor basis. For the mixing angle we obtain $\phi^\phys=39.3(2.0)^\circ$ and our results for the decay-constant parameters are given by $f_l^\phys=138.6(4.4)\mev$ and $f_s^\phys=170.7(3.3)\mev$.

\end{abstract}

\maketitle
\clearpage

\section{Introduction}
\label{sec:introduction}

The standard model (SM) of particle physics is largely defined via its
symmetries. In particular, the invariance under local gauge
transformations gives rise to the gauge bosons acting as force
carriers for electromagnetic, weak, and strong interactions.
The strongly interacting part of the SM, quantum chromodynamics
(QCD), exhibits peculiarly broken symmetries. First, there
is chiral symmetry, which is spontaneously broken and, thereby,
leading to an octet of pseudo-scalar mesons playing the role of
Goldstone bosons, which are massless in the absence of an explicit
quark-mass term in the corresponding action. Second, there is the
anomalously broken axial U$(1)$ symmetry thought to be responsible for
the large mass of a flavor-singlet pseudo-scalar meson. In nature we
find the pion triplet and four kaons, representing seven of the eight
octet states. The eighth would be the $\eta$, which is, however,
mixing with the singlet state, the $\eta^\prime$, due to SU$(3)$
flavor-breaking effects from finite quark-mass values.

The investigation of $\eta$ and $\eta^\prime$ mesons
and related quantities is not only interesting for our understanding 
of symmetry-breaking mechanisms in QCD. It is also highly relevant for
phenomenology, like for instance the anomalous magnetic moment of the
muon, where $\eta$ and $\eta^\prime$ mesons play an important role for the
hadronic light-by-light contribution (for recent progress on this see
Refs.~\cite{ExtendedTwistedMass:2022ofm,Gerardin:2023naa,Holz:2024diw,Holz:2024lom,Estrada:2024cfy}
and references therein).
$\eta$ and $\eta^\prime$ also play fundamental roles in the famous
Witten-Veneziano formula~\cite{Witten:1979vv,Veneziano:1979ec,Cichy:2015jra,Cichy:2015svi},
and for the coupling to axions or axion-like dark matter candidates~\cite{Gao:2022xqz}.
For a recent review on precision tests of fundamental physics see Ref.~\cite{Gan:2020aco}.

The theoretical investigation of $\eta, \eta^\prime$ physics requires
a non-perturbative method, with the method of choice given by lattice
QCD (LQCD), where Monte Carlo simulations are performed in 
discretized Euclidean space-time. For a long time technical reasons prevented
simulations with quark-mass values tuned such that, e.g., physical pion,
kaon, and $D$-meson masses are all reproduced. However, due to
algorithmic advances such simulations have become standard now.

Still, most of the previous LQCD investigations of $\eta,\eta^\prime$
physics were based on ensembles with unphysical quark-mass values,
requiring an extrapolation to the physical point. For $N_f=2$
dynamical quark flavors, results can be found in
Ref.~\cite{Jansen:2008wv}, for $N_f=2+1$ or $N_f=2+1+1$ in
Refs.~\cite{Ottnad:2012fv,Michael:2013gka,Cichy:2015jra,Ottnad:2017bjt,Bali:2021qem}. A 
first exploratory calculation with $N_f=2$ dynamical quark flavors at
the physical point with Wilson clover twisted-mass fermions can be
found in Ref.~\cite{Dimopoulos:2018xkm}, and a more recent
calculation with $N_f=2+1$ dynamical rooted staggered quarks including
physical-point ensembles in Ref.~\cite{Verplanke:2024msi}. 

In this paper we investigate $\eta$- and $\eta^\prime$-meson masses,
the mixing angle as well as decay constants and related quantities
based on $N_f=2+1+1$ gauge field configurations of the Extended Twisted
Mass Collaboration
(ETMC)~\cite{Alexandrou:2018egz,ExtendedTwistedMass:2021qui,Finkenrath:2022eon} 
(see also Ref.~\cite{ExtendedTwistedMass:2022jpw}). This includes in particular three ensembles
with average up/down-, strange- and charm-quark masses tuned to their
physical values~\cite{Alexandrou:2018egz}. These three ensembles cover three values of
the lattice spacing down to $a=0.057\ \mathrm{fm}$. In addition, we include ensembles
with larger-than-physical pion-mass values, and also at a fourth lattice
spacing value in the analysis. The lattice formulation, the Wilson clover twisted-mass
formulation at maximal twist has the important property of automatic
$O(a)$ improvement. This allows us to
perform a controlled continuum extrapolation while avoiding an
additional extrapolation in the light-quark mass values. As a result we are able to reduce uncertainties significantly compared
to Ref.~\cite{Ottnad:2017bjt}.

\section{Lattice setup} \label{sec:lattice_setup}
The lattice calculations for this study are carried out on 13 gauge ensembles that have been generated by the Extended Twisted Mass Collaboration (ETMC) with $N_f=2+1+1$ flavors of twisted-mass clover-improved fermions~\cite{Alexandrou:2018egz,ExtendedTwistedMass:2020tvp,ExtendedTwistedMass:2021qui,Finkenrath:2022eon,Sheikholeslami:1985ij} and the Iwasaki-improved gauge action \cite{Iwasaki:1985we}. The ensembles have been tuned to maximal twist to achieve automatic $O(a)$ improvement for the computation of physical observables~\cite{Frezzotti:2003ni,Frezzotti:2004wz,Frezzotti:2005gi}. A summary of the ensembles is given in Table~\ref{tab:ensembles}, including the bare light-quark masses and corresponding values of $M_\pi$. The ensembles cover four values of the lattice spacing and include three ensembles at physical quark masses, as well as three more ensembles with $M_\pi<200\mev$. The ensembles in Table~\ref{tab:ensembles} have been generated with a mass-degenerate doublet of light quarks and a non-degenerate doublet of heavy (strange and charm) quarks. The corresponding Dirac operators employed in the sea part of the action read
\begin{equation}
  D_l(U) = D_W(U) + \frac{i}{4} c_\mathrm{SW} \sigma^{\mu\nu} F^{\mu\nu}(U) + m_\mathrm{crit} + i\mu_l \tau^3 \g{5} \,,
  \label{eq:D_l}
\end{equation}
and
\begin{equation}
  D_h(U) = D_W(U) + \frac{i}{4} c_\mathrm{SW} \sigma^{\mu\nu} F^{\mu\nu}(U) + m_\mathrm{crit} - \mu_\delta \tau^1 + i\mu_\sigma\tau^3\g{5} \,,
  \label{eq:D_h}
\end{equation}
respectively, where $D_W$ denotes the massless Wilson Dirac operator, $\mu_{l,\sigma,\delta}$ bare quark mass parameters and $m_\mathrm{crit}$ the critical value of the bare Wilson quark mass. The clover term $\frac{i}{4} c_\mathrm{SW} \sigma^{\mu\nu} F^{\mu\nu}(U)$ has been included to suppress $\mathcal{O}(a^2)$ lattice artifacts in the neutral-pion mass~\cite{Becirevic:2006ii,Dimopoulos:2009es,Herdoiza:2013sla,ETM:2015ned,Finkenrath:2017cau} that might otherwise render the simulation unstable at coarser values of the lattice spacing. The value of the Sheikoleslami-Wohlert improvement coefficient $c_\mathrm{SW}$~\cite{Sheikholeslami:1985ij} has been determined using 1-loop tadpole-boosted perturbation theory~\cite{Aoki:1998qd}. The sea strange- and charm-quark masses have been tuned to their physical values on all of these ensembles by the procedure detailed in Refs.~\cite{Alexandrou:2018egz,ExtendedTwistedMass:2020tvp}.\\
We note in passing that for ensemble cA211.12.48 (see Table~\ref{tab:ensembles}) reweighting was required to achieve the condition for maximal twist.


\begin{table}[!t]
 \caption{List of $N_f=2+1+1$ gauge ensembles that have been used in this work. Conversion to physical units for $M_\pi$ and $L$ has been carried out using the scale-setting procedure described in the text. $N_\mathrm{conf}$ denotes the number of gauge configurations on which observables including $M_\pi$ have been measured.}
 \centering
 \setlength{\tabcolsep}{0.5em}
 \begin{tabular}{lcrccccr}
  \hline\hline
  ID           & $\beta$ & $\frac{T}{a}\times\bigl(\frac{L}{a}\bigr)^3$ & $a \mu_l$ & $\Mpi/\mev$ & $\Mpi L$ & $L/\fm$ & $N_\mathrm{conf}$  \\
  \hline\hline                                                                      
  cA211.12.48  & 1.726 &  $96 \times 48^3$ & 0.00120 & 171  & 3.85 & 4.42 &  287 \\ 
  cA211.15.48  &       &  $96 \times 48^3$ & 0.00150 & 191  & 4.27 & 4.42 & 1853 \\ 
  cA211.30.32  &       &  $64 \times 32^3$ & 0.00300 & 268  & 4.01 & 2.95 & 1261 \\ 
  cA211.40.24  &       &  $48 \times 24^3$ & 0.00400 & 311  & 3.48 & 2.21 & 1320 \\ 
  cA211.53.24  &       &  $48 \times 24^3$ & 0.00530 & 356  & 4.00 & 2.21 &  624 \\ 
  \hline
  cB211.072.64 & 1.778 & $128 \times 64^3$ & 0.00072 & 139  & 3.62 & 5.12 &  779 \\ 
  cB211.14.64  &       & $128 \times 64^3$ & 0.00140 & 193  & 5.01 & 5.12 &  456 \\ 
  cB211.25.48  &       &  $96 \times 48^3$ & 0.00250 & 257  & 5.01 & 3.84 &  654 \\ 
  cB211.25.32  &       &  $64 \times 32^3$ & 0.00250 & 260  & 3.37 & 2.56 &  989 \\ 
  cB211.25.24  &       &  $48 \times 24^3$ & 0.00250 & 271  & 2.64 & 1.92 &  574 \\ 
  \hline
  cC211.06.80  & 1.836 & $160 \times 80^3$ & 0.00060 & 137  & 3.80 & 5.47 &  738 \\ 
  cC211.20.48  &       &  $96 \times 48^3$ & 0.00200 & 248  & 4.12 & 3.28 &  611 \\ 
  \hline
  cD211.054.96 & 1.900 & $192 \times 96^3$ & 0.00054 & 140  & 3.89 & 5.50 &  492 \\ 
  \hline\hline
 \end{tabular}
 \label{tab:ensembles}
\end{table}

Throughout the analysis, dimensionful observables are expressed in units of the gradient-flow scale $t_0$ introduced in Ref.~\cite{Luscher:2010iy}, using the values for $\sqrt{t_0}/a$ in Table~\ref{tab:beta_dependent_parameters}. The values of $\sqrt{t_0}/a$ at physical quark mass for the three coarsest lattice spacings have already been determined in Ref.~\cite{ExtendedTwistedMass:2021qui} and the value for the finest lattice spacing has been computed on the cD211.054.96 ensemble. For the scale setting we use the physical value of $\sqrt{t_0}$ for $N_f=2+1+1$ dynamical quark flavors that has been reported in Ref.~\cite{ExtendedTwistedMass:2021qui}, i.e.,
\begin{equation}
 \sqrt{t_0^\phys} = 0.14436(61) \fm .
 \label{eq:sqrt_t0_phys}
\end{equation}
The values for $M_\pi$ and $L$ in physical units as listed in Table~\ref{tab:ensembles} have been obtained by the very same procedure.


\begin{table}[!t]
 \caption{Additional, $\beta$-dependent parameters. The values of the gradient-flow scale $\sqrt{t_0}/a$ at physical quark mass for the three smallest value of $\beta$ are taken from Ref.~\cite{ExtendedTwistedMass:2021qui}, whereas the value at $\beta=1.900$ has been computed for this work. The corresponding lattice spacings $a$ have been determined using the physical value of $\sqrt{t_0}$ in Eq.~(\ref{eq:sqrt_t0_phys}). The values for the ratio of the (non-singlet) pseudoscalar and scalar renormalization factors $Z_P/Z_S$ have been computed in Ref.~\cite{ExtendedTwistedMass:2022jpw} from hadronic observables at physical pion mass. The comparatively larger error on $Z_P/Z_S$ at $\beta=1.726$ is due to a chiral extrapolation as there is no ensemble with physical quark mass available at the coarsest lattice spacing. The three values of the valence strange-quark mass $\mu^\mathrm{val,i}_s$, $i=1,2,3$ are used in the calculation of primary observables and are different from the three target values $\mu_s^{M1,\ldots,M3}$ obtained from the matching conditions in Eqs.~(\ref{eq:M1})--(\ref{eq:M3}), cf. Subsection~\ref{subsec:mu_s_matching}.}
 \centering
 \setlength{\tabcolsep}{0.5em}
 \begin{tabular}{cccccccccc}
  \hline\hline
  $\beta$ & $\sqrt{t_0}/a$ & $a/\fm$ & $Z_P/Z_S$ & $\mu^\mathrm{val,1}_s$ & $\mu^\mathrm{val,2}_s$ & $\mu^\mathrm{val,3}_s$ & $\mu_s^{M1}$ & $\mu_s^{M2}$ & $\mu_s^{M3}$ \\
  \hline\hline
    1.726 & 1.5660(22)  & 0.0922(4) & 0.75171(288) & 0.0176 & 0.0220 & 0.0264 & 0.01817(61) & 0.020716(95) & 0.016856(54) \\
    1.778 & 1.80396(68) & 0.0800(3) & 0.79066(23)  & 0.0148 & 0.0185 & 0.0222 & 0.01659(17) & 0.018471(62) & 0.016355(13) \\
    1.836 & 2.1094(8)   & 0.0684(3) & 0.82308(23)  & 0.0128 & 0.0161 & 0.0193 & 0.01510(21) & 0.016265(74) & 0.015106(10) \\
    1.900 & 2.51821(76) & 0.0573(2) & 0.85095(18)  & 0.0136 & 0.0140 & 0.0150 & 0.01323(19) & 0.013768(43) & 0.013327(08) \\
  \hline\hline
 \end{tabular}
 \label{tab:beta_dependent_parameters}
\end{table}

Statistical errors on observables are generally computed and propagated through non-parametric bootstrapping. To this end, we carry out the entire analysis on $N_B=10000$ bootstrap samples while using binning to account for autocorrelation in the data. Errors on additional input parameters are included via a parametric bootstrapping procedure. The final error estimates including systematics involve model averages and a separation of statistical and systematic errors which will be discussed in Sec.~\ref{sec:AIC_final_results}.

\subsection{Valence action} \label{subsec:valence_action}
The mass-splitting term $\sim \mu_\sigma \tau^1$ for the heavy doublet in Eq.~(\ref{eq:D_h}) introduces mixing between strange- and charm-quark flavors at finite values of the lattice spacing. While it is possible to deal with the resulting complications for the computation of observables related to $\eta$ and $\etap$ as demonstrated in Refs.~\cite{Ottnad:2012fv,Michael:2013gka,Ottnad:2017bjt}, they can be readily avoided by switching to a mixed-action setup for the heavy-quark sector. To this end, we employ Osterwalder-Seiler (OS) fermions \cite{Osterwalder:1977pc} in the discretization of heavy-quark flavors. Moreover, as argued in Refs.~\cite{Ottnad:2012fv,Michael:2013gka,Ottnad:2017bjt} the charm-quark contribution to the $\eta$--$\eta'$ system can be safely neglected, hence we only consider the strange-quark contribution. In this approach the strange-quark contribution is implemented by introducing two strange-quark flavors $s$ and $s'$ with bare mass $\mu_s^\mathrm{val} = -\mu_{s'}^\mathrm{val}$. The lattice action is extended by a fermionic (valence) action based on the Dirac operator for the two flavors $s$ and $s'$
\begin{equation}
 D_{s,s'}(U) = D_W(U) + \frac{i}{4} c_\mathrm{SW} \sigma^{\mu\nu} F^{\mu\nu}(U) + m_\mathrm{crit} + i\mu_{s,s'}^\mathrm{val} \g{5} \,.
 \label{eq:D_OS}
\end{equation}
Following Ref.~\cite{Frezzotti:2004wz}, the valence action is complemented by a ghost action to cancel unphysical contributions from the additional valence quarks to the fermionic determinant, reproducing the correct continuum limit. Since Eq.~(\ref{eq:D_OS}) is diagonal in flavor space the flavor mixing pertinent to the unitary setup is avoided in the valence sector. Furthermore, this choice of the valence action preserves automatic $O(a)$ improvement \cite{Frezzotti:2004wz} and allows us to apply a more efficient variance reduction technique in the computation of quark-disconnected diagrams involving strange quarks \cite{Ottnad:2015hva}. For an earlier account of this approach in the context of $\eta$ and $\etap$ mesons, as well as further, technical details we refer to the exploratory study in Ref.~\cite{Ottnad:2015hva}. \par

The introduction of OS valence quarks $s$ and $s'$ necessitates fixing the bare valence quark-mass parameter $\mu_s^\mathrm{val}$. While the sea-action parameters $\mu_\sigma$ and $\mu_\delta$ have been tuned to reproduce the physical strange (and charm) quark mass(es), the values of $\mu_s^\mathrm{val}$ generally depend on the matching condition imposed to reproduce the correct strange-quark mass in the continuum limit. One possible choice to determine $\mu_s^\mathrm{val}$ is a direct matching with the bare strange-quark mass $\mu_s$ in the unitary setup, i.e., imposing
\begin{equation}
 \mu_s^\mathrm{val} \equiv \mu_s = \mu_\sigma - \frac{Z_P}{Z_S} \mu_\delta \,.
 \label{eq:mu_s}
\end{equation}
However, as already pointed out in Ref.~\cite{Ottnad:2015hva}, the resulting value of $\mu_s^\mathrm{val}$ is very sensitive to any uncertainty in the ratio of flavor non-singlet pseudoscalar and scalar renormalization factors $Z_P/Z_S$, which is why we refrain from implementing this direct way of matching quark masses but rather use matching conditions based on hadron masses. Since the target values for $\mu_s^\mathrm{val}$ that result from such a matching condition are not known \emph{a priori}, we have carried out the valence part of our lattice simulations for a suitable range of three distinct values $\mu_s^{\mathrm{val},i}$, $i=1,2,3$ as listed in Table~\ref{tab:beta_dependent_parameters} for each value of $\beta$. This allows us to map out the dependence of any observable on $\mu_s^\mathrm{val}$ such that an interpolation (extrapolation) to the final target value from a given matching condition can be carried out at sufficiently high accuracy. The details of the matching procedure will be discussed in Subsec.~\ref{subsec:mu_s_matching}.

\subsection{Correlation Functions} \label{subsec:correlation_functions}
The computation of masses and amplitudes for mesons proceeds through the evaluation of standard Euclidean two-point correlation functions at vanishing momentum
\begin{equation}
 C_{\mathcal{O}_1\mathcal{O}_2}(t) = \langle \mathcal{O}_1(t+t_i) \mathcal{O}_2^\dag(t_i) \rangle
 \label{eq:2pt}
\end{equation}
where $\mathcal{O}_i(t)$ are suitable interpolating operators for the desired meson(s) and $t_i$ denotes the source timeslice. For the pion we consider the standard, pseudoscalar interpolating operator in the twisted basis, i.e.,
\begin{equation}
 \mathcal{O}^{\pm}_\pi(t) = \frac{1}{\sqrt{2}} \sum_{\vec{x}} \bar{\chi}_l(x) i\g{5}\tau^\pm \chi_l(x)\,,
 \label{eq:O_pi_pm_tm}
\end{equation}
where the sum is over spatial coordinates $\vec{x}$ at fixed $t$, $\chi_l=(u,d)^T$ and following Ref.~\cite{EuropeanTwistedMass:2010voq} we have defined the projector $\tau^\pm=\frac{\tau^1 \pm i\tau^2}{2}$. For the kaon involving the Osterwalder-Seiler strange quark $s$, we use 
\begin{equation}
 \mathcal{O}_K(t) = \sum_{\vec{x}} \bar{s}(x) i\g{5} u(x) \,.
 \label{eq:O_OS_K}
\end{equation}
As discussed in Ref.~\cite{Ottnad:2015hva}, this choice is not unique but differs by (potentially large) lattice artifacts from the ``neutral'' kaon obtained from using $\mathcal{O}^{\mathrm{OS}}_{K^0} = \sum_\vec{x} \bar{s}'(x) i\g{5} d(x)$. The interpolating operators for the $\eta$ and $\etap$ are given by
\begin{align}
 \mathcal{O}_l(t) &= -\frac{1}{\sqrt{2}} \sum_{\vec{x}} \bar{\chi}_l(x) \tau^3 \chi_l(x) \,, \label{eq:O_OS_l} \\
 \mathcal{O}_s(t) &= -\sum_{\vec{x}} \bar{s}(x) s(x) \,, \label{eq:O_OS_s} 
\end{align}
where we have chosen the flavor structure according to the quark-flavor basis. Note that the axial rotation from the physical to the twisted-mass basis \cite{Frezzotti:2000nk,Frezzotti:2004wz} has already been carried out. It is responsible for turning pseudoscalar operators in the physical basis into scalar operators in the twisted basis \cite{Shindler:2007vp}. However, unlike in the unitary setup there is no mixing between scalar and pseudoscalar operators for heavy quarks, cf.~Refs.~\cite{Ottnad:2012fv,Michael:2013gka,Ottnad:2017bjt}. From this set of operators we build a $2\times2$ correlation function matrix
\begin{equation}
 \mathcal{C}(t) = \l(\begin{array}{cc}
  C_{ll}(t) & C_{ls}(t) \\
  C_{sl}(t) & C_{ss}(t)
  \end{array}\r) \,,
 \label{eq:corr_matrix}
\end{equation}
where we have introduced the shorthands $C_{ij}(t) = C_{\mathcal{O}_i\mathcal{O}_j}(t)$ for $i,j=l,s$. The diagonal elements consist of a quark-connected and a quark-disconnected contributions, whereas the off-diagonal elements $C_{ls}(t)=C_{sl}(t)$ that convey the physical mixing between flavor eigenstates are given purely by quark-disconnected diagrams. \par

\begin{figure}[t]
 \centering
  \includegraphics[totalheight=0.226\textheight]{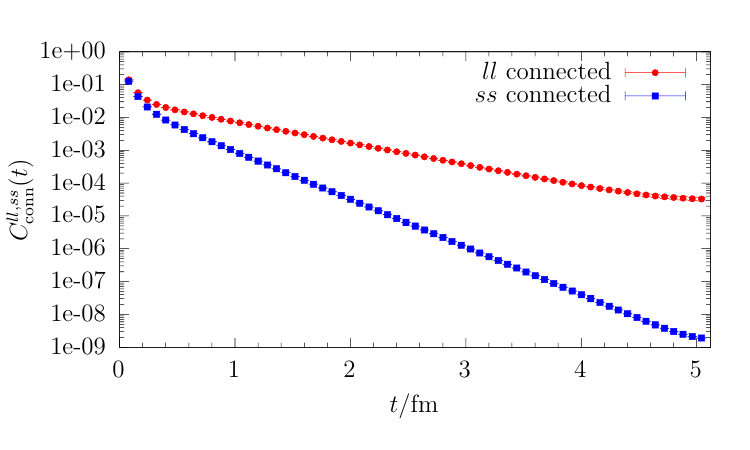}
  \includegraphics[totalheight=0.226\textheight]{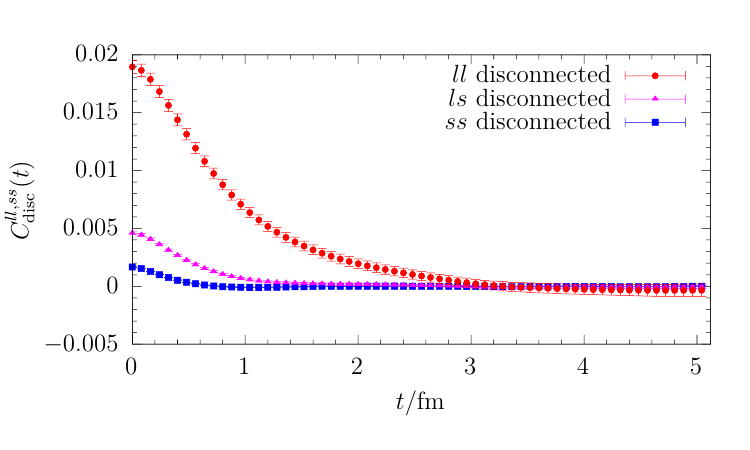}
  \caption{Quark-connected (left panel) and quark-disconnected (right panel) two-point correlation functions for light and strange quarks on the cB211.072.64 ensemble.}
  \label{fig:correlators}
\end{figure}

For the computation of quark-connected diagrams we use stochastic timeslice sources and the one-end trick \cite{ETM:2008zte,Foster:1998vw,McNeile:2006bz}. We generally average over multiple source positions $t_i$ to further improve the signal quality. The timeslices $t_i$ are randomly distributed as all our lattice boxes have been generated with periodic boundary conditions in time. On the other hand, the quark-disconnected pieces contributing to the $\eta$--$\etap$ system are computed using undiluted stochastic volume sources and the one-end trick. Unlike in the unitary setup, the one-end trick can be used for both, light and strange quarks as shown in Ref.~\cite{Ottnad:2015hva}. Together with using $\mathcal{O}(100)$ stochastic volume sources, this guarantees the signal to be dominated by gauge noise on every single ensemble. Fig.~\ref{fig:correlators} shows examples for the signal quality that is obtained for the quark-connected and disconnected contributions to the $\eta$ and $\etap$ correlation function matrix in Eq.~(\ref{eq:corr_matrix}) at physical quark masses. \par

Furthermore, we require correlation functions for the artificial $\eta_s$ meson and the $\Omega$ baryon to compute their respective masses. These are used in matching conditions to determine target values for $\mu_s^\mathrm{val}$. Following Ref.~\cite{Ottnad:2015hva}, the corresponding interpolating operator for the $\eta_s$ can be defined as
\begin{equation}
 \mathcal{O}_{\eta_s}(t) = -\sum_{\vec{x}} \bar{s}(x) s''(x) \,,
 \label{eq:O_OS_eta_s}
\end{equation}
where yet another Osterwalder-Seiler strange quark $s''$ with $\mu_s=\mu_{s''}$ has been introduced to ensure the absence of quark-disconnected diagrams in the corresponding two-point function. From a practical point of view, this amounts to computing only the quark-connected part of the pure strange contributions to the $\eta$, $\etap$ correlation function matrix, i.e., $C_{\eta_s}(t) = C^\mathrm{conn}_{ss}(t)$. 

For the $\Omega$ baryon we compute zero-momentum $\Omega$ baryon correlation functions following the conventions in Ref.~\cite{Alexandrou:2014sha},
\begin{align}
C^{\pm}_{\Omega\Omega}(t)=\sum_{\vec{x}_{\mathrm{sink}}-\vec{x}_{\mathrm{source}}}\l\langle \trace{\Gamma_\pm {\mathcal O}_{\Omega}(\vec{x}_{\mathrm{sink}},t+t_i){\mathcal O}_{\Omega}^{\dagger}(\vec{x}_{\mathrm{source}},t_i)}\r\rangle\,,
\end{align} 
where ${\mathcal O_\Omega}=\sum_{i\in{1,2,3}}\varepsilon_{abc}\left(s^a C\gamma_is^{b}\right)s^c(\vec{x},t)$ is the interpolating field for the omega baryon on a point-source at $(\vec{x},t)$, and $\Gamma_\pm = \frac{1}{4}\left(1\pm \gamma_0\right)$ spin-projects onto states of positive and negative parity. The forward part of the positive parity correlation functions is connected to the 
backward part of the negative parity correlation function due to discrete symmetries of the twisted-mass formulations, therefore we average correlators forward, backward in time:
\begin{align}
C_{\Omega\Omega}(t)=C^{+}_{\Omega\Omega}(t)-C^{-}_{\Omega\Omega}(T-t).
\end{align}
In this work we used $\mathcal{O}(100)$ source positions per gauge configuration.

\section{Observables and methods} \label{sec:observables_and_methods}
The spectral decompositions of the mesonic two-point functions defined in the previous section all have the usual form
\begin{equation}
 C_{ij}(t) = \sum_{n} \frac{A^n_i (A^n_j)^*}{2E_n} \l( e^{-M_n t} + e^{M_n(t-T)} \r) \,,
 \label{eq:spectral_decomposition}
\end{equation}
where $i$, $j$ label the various interpolating operators at sink and source and $n$ is used to index physical states. The (unrenormalized) amplitudes $A_{i,j}^n$ are related to decay constants and mixing parameters and $M_n$ denotes the masses of the corresponding state. In general, the decay constants $f_a^P$ of pseudoscalar mesons $P=\pi,K,\eta,\etap,\ldots$ with respect to some flavor structure $a$ are defined from axial-vector matrix elements
\begin{equation}
 \bra{0} \mathcal{A}^\mu_a \ket{P(p)} = i f_a^P p^\mu \,.
 \label{eq:decay_constants}
\end{equation}
However, in the twisted basis the pion decay constant $f_\pi$ can be readily computed from pseudoscalar matrix elements as obtained from two-point functions based on the interpolating operators in Eq.~(\ref{eq:O_pi_pm_tm})
\begin{equation}
 f_\pi = 2\mu_l \frac{\bra{0} \mathcal{O}^{\pm}_\pi \ket{\pi}}{M_\pi^2}
 \label{eq:f_pi}
\end{equation}
This is possible because the chiral rotation turns the axial-vector current in the physical basis into a vector current in the twisted basis at maximal twist, which can be traded for the pseudoscalar current through the usual relation for the partially conserved vector current \cite{Frezzotti:2001du,DellaMorte:2001tu,Jansen:2003ir}. Note that this expression does not require any renormalization. Similarly, the kaon decay constant can be obtained from pseudoscalar matrix elements as well, i.e.,
\begin{equation}
 f_K = \l(\mu_l + \mu_s\r) \frac{\bra{0} \mathcal{O}_K \ket{K}}{M_K^2} \,.
 \label{eq:f_K}
\end{equation}
Again there is no need for renormalization at the operator level. \par 

Restricting ourselves to light and strange valence quarks, Eq.~(\ref{eq:decay_constants}) implies a $2\times2$ matrix for the mixing of flavor to physical eigenstates for the $\eta$ and $\etap$ mesons, that can be parametrized as
\begin{equation}
 \l(\begin{array}{cc}
    f_a^\eta    & f_b^{\eta} \\
    f_a^{\etap} & f_b^{\etap}
    \end{array}\r) = 
 \l(\begin{array}{rr}
    f_a \cos\phi_a & -f_b \sin\phi_b \\
    f_a \sin\phi_a &  f_b \cos\phi_b
    \end{array}\r)\,,
 \label{eq:mixing}
\end{equation}
where $a\neq b$ refer to the flavor basis for the two axial-vector currents $\mathcal{A}^\mu_{a,b}(x)$. There exist two popular choices of basis in the literature, namely the octet-singlet basis ($a=8$, $b=0$)
\begin{align}
 \mathcal{A}^\mu_8(x) &= \frac{1}{\sqrt{6}}\l(\bar{u}(x)\g{\mu}\g{5}u(x) + \bar{d}(x)\g{\mu}\g{5}d(x) - 2\bar{s}(x)\g{\mu}\g{5}s(x)\r) \,, \label{eq:A_8_mu} \\
 \mathcal{A}^\mu_0(x) &= \frac{1}{\sqrt{3}}\l(\bar{u}(x)\g{\mu}\g{5}u(x) + \bar{d}(x)\g{\mu}\g{5}d(x) +  \bar{s}(x)\g{\mu}\g{5}s(x)\r) \,, \label{eq:A_0_mu}
\end{align}
and the quark-flavor basis ($a=l$, $b=s$)
\begin{align}
 \mathcal{A}^\mu_l(x) &= \frac{1}{\sqrt{2}}\l(\bar{u}(x)\g{\mu}\g{5}u(x) + \bar{d}(x)\g{\mu}\g{5}d(x) \r) \,, \label{eq:A_l_mu} \\
 \mathcal{A}^\mu_s(x) &= \bar{s}(x) \g{\mu}\g{5}s(x)\,. \label{eq:A_s_mu}
\end{align}
These two bases are special in the sense that in chiral perturbation theory ($\chi$PT) the leading-order difference of the mixing angles $\l|\phi_a-\phi_b\r|$ is entirely due to a SU(3) flavor-breaking contribution in the octet-singlet basis, whereas in the quark-flavor basis it is given by an Okubo-Zweig-Iizuka (OZI) violating term. Moreover, in the
SU(3) flavor-symmetric theory one finds $\phi_0=\phi_8=0$ and $\phi_l=\phi_s=\arctan\sqrt{2}$, respectively. The resulting smallness of the difference in the mixing angles in the quark-flavor basis
\begin{equation}
 \l|\frac{\phi_l-\phi_s}{\phi_l+\phi_s}\r| \ll 1 
 \label{eq:FKS_motivation}
\end{equation}
motivates the so-called Feldmann-Kroll-Stech (FKS) scheme \cite{Feldmann:1998vh,Feldmann:1998sh} that includes only a single mixing angle in the quark-flavor basis
\begin{equation}
 \l(\begin{array}{cc}
    f_l^\eta    & f_s^{\eta} \\
    f_l^{\etap} & f_s^{\etap}
    \end{array}\r) = 
 \l(\begin{array}{rr}
    f_l \cos\phi & -f_s \sin\phi \\
    f_l \sin\phi &  f_s \cos\phi
    \end{array}\r)\,.
 \label{eq:FKS_scheme}
\end{equation}

In this particular mixing scheme the decay constant parameters $f_l$ and $f_s$ as well as the mixing angle $\phi$ do not exhibit scale dependent contributions as they formally only enter at higher order \cite{Kaiser:1998ds,Feldmann:1998vh}. For further details on this subject we refer to the comprehensive review in Ref.~\cite{Feldmann:1999uf}. Following the same strategy as in Refs.~\cite{Michael:2013gka,Ottnad:2017bjt} our lattice calculations are carried out in the quark-flavor basis employing the FKS scheme. This allows us to compute $f_l$, $f_s$ and $\phi$ from (renormalized) pseudoscalar matrix elements 
\begin{equation}
 h_{l,s}^P = 2 \mu_{l,s} \frac{Z_S}{Z_P} \bra{0} \mathcal{O}_{l,s} \ket{P} \,, \quad P=\eta,\etap 
 \label{eq:hfP}
\end{equation}
that are extracted from the correlation functions based on the interpolating operators in Eqs.~(\ref{eq:O_OS_l})~and~(\ref{eq:O_OS_s}) and by making use of the relation

\begin{equation}
 \l(\begin{array}{cc}
   h_l^\eta    & h_s^{\eta} \\
   h_l^{\etap} & h_s^{\etap}
   \end{array}\r) =
 \l(\begin{array}{rr}
   \cos\phi & -\sin\phi \\
   \sin\phi &  \cos\phi
 \end{array}\r)\diag\l(M_\pi^2 f_l,\, (2M_K^2-M_\pi^2) f_s\r)\,,
 \label{eq:PS_mixing}
\end{equation}
which holds to the same order in $\chi$PT that has been used for the splitting of mixing angles in the FKS scheme \cite{Feldmann:1999uf}. \par

\subsection{Extraction of masses and amplitudes} \label{subsec:extraction_of_masses and amplitudes}
The required masses and amplitudes are extracted from lattice data for two-point functions that are built from the various interpolating operators defined in subsection~\ref{subsec:correlation_functions}. In case of the pion, kaon and the $\eta_s$ meson only quark-connected diagrams contribute that are not affected by an exponential decay of the signal-to-noise ratio. Therefore, the groundstate contribution in Eq.~(\ref{eq:spectral_decomposition}) can be readily extracted from a fit at sufficiently large Euclidean time separations $t/a$. The corresponding results are collected in Table~\ref{tab:M_pi_M_K_and_M_eta_s}. On the other hand, for the $\eta$ and $\etap$ states masses and amplitudes can be extracted by solving a generalized eigenvalue problem (GEVP)~\cite{Michael:1982gb,Luscher:1990ck,Blossier:2009kd,Fischer:2020bgv,Ostmeyer:2024qgu} for the correlation function matrix $\mathcal{C}(t)$
\begin{equation}
 \mathcal{C}(t) v^n(t,t_0) = \lambda^n(t, t_0) \mathcal{C}(t_0) v^n(t,t_0) \,,
 \label{eq:GEVP}
\end{equation}
and fitting the eigenvalues (principal correlators) $\lambda(t)$ and eigenvectors $v$ at sufficiently large values of $t/a$. However, due to the severe signal-to-noise problem caused by the quark-disconnected contributions additional steps are required to improve the signal, similar to what has been used in Ref.~\cite{Ottnad:2017bjt}, and they are briefly summarized in the following. \par

In a first step, we replace the quark-connected pieces in $C_{ll}(t)$ and $C_{ss}(t)$ that do not suffer from a signal-to-noise problem by their respective groundstate contributions, thus effectively removing any excited-state contamination stemming from quark-connected pieces. The relevant groundstate contribution are reconstructed from the parameters obtained from a simple, correlated fit at large values of $t/a$ for each bootstrap sample. This kind of replacement originally introduced in Ref.~\cite{Neff:2001zr} has been shown in previous studies \cite{Jansen:2008wv,Michael:2013gka,Ottnad:2017bjt} to greatly reduce the excited-state contamination in the resulting $\etap$ and $\etap$ states, thus facilitating fits at much smaller values of $t/a$ before the signal is lost in noise. \par

\begin{figure}[t]
 \centering
 \includegraphics[totalheight=0.226\textheight]{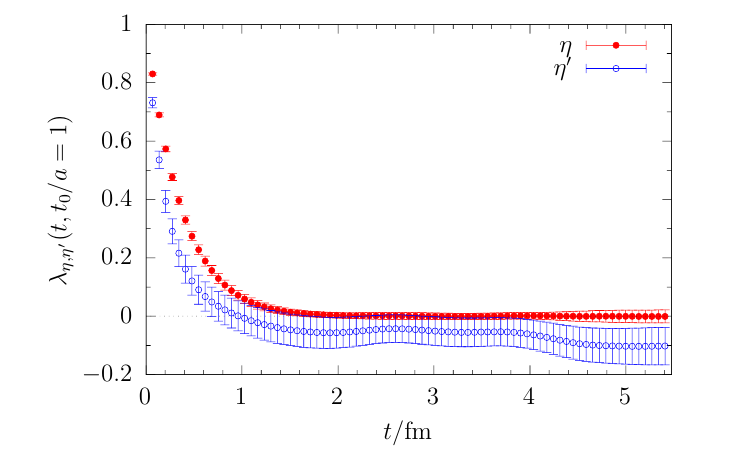}
 \includegraphics[totalheight=0.226\textheight]{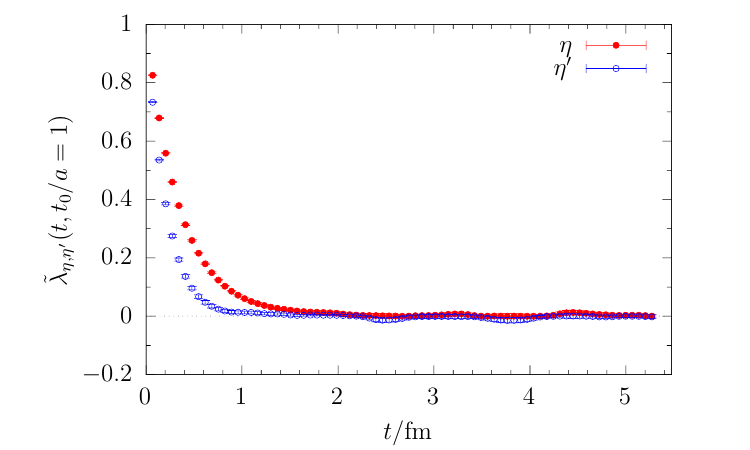}
 \caption{Effect of finite topological sampling on the $\eta$ and $\etap$ principal correlators at physical quark mass on the cC211.06.80 ensemble. Data are shown for solving the GEVP on the unaltered correlation functions (left panel) and on the second discrete time-derivative of the correlation functions (right panel).}
 \label{fig:topological_FV_effect}
\end{figure}

In a second step, we improve the signal with respect to the effects of finite topological sampling. As shown in Ref.~\cite{Aoki:2007ka}, the quark-disconnected part of a pseudoscalar flavor-singlet correlation functions evaluated in finite volume and finite topological sampling receives a non-trivial constant offset at large Euclidean time separations, which for (exactly) fixed topology takes the form
\begin{equation}
 \sim \frac{1}{V} \l(\chi_t -\frac{Q^2}{V} + \frac{c_4}{2V\chi_t}\r) + \mathcal{O}\l(V^{-2}\r) \,,
\end{equation}
where $\chi_t$ and $c_4$ refer to the 2nd and 4th moment of the topological charge distribution, i.e., the topological susceptibility and kurtosis of the distribution. This contribution only vanishes as $V\rightarrow\infty$ or for perfect sampling over all topological sectors, i.e., in the limit of infinite gauge statistics. The resulting constant offset in the correlation functions for the $\eta$ and $\etap$ is typically found to be very noisy and strongly affected by autocorrelation. It is observed in our lattice data particularly for ensembles at smaller values of $M_\pi$ and finer lattice spacings, as expected due to the slowdown of topological sampling. While such a constant shift could in principle be fitted \cite{Bali:2014pva}, we prefer to remove it by taking discrete time-derivatives of the correlation functions as introduced in Ref.~\cite{Ottnad:2017bjt} for this purpose. Removing the constant shift by a discrete derivative in $t$  has the added advantage of greatly reducing the autocorrelation in the final signal as well as data correlation in $t$. Therefore, less data binning is required in the bootstrapping procedure, and correlated fits become more stable. For our set of data we empirically find that taking the second derivative further decorrelates the data in $t$, while even higher derivatives rapidly deteriorate the overall signal-to-noise behavior. This motivates the replacement
\begin{equation}
 \mathcal{C}(t) \rightarrow \tilde{\mathcal{C}}(t) = \mathcal{C}(t) - 2\mathcal{C}(t+1) + \mathcal{C}(t+2)
 \label{eq:corr_function_derivative}
\end{equation}
before solving the GEVP in Eq.~(\ref{eq:GEVP}). Figure \ref{fig:topological_FV_effect} shows an example for the efficacy of this method on the cC211.06.80 ensemble at physical quark masses, comparing results for the eigenvalues $\lambda^{\eta,\etap}(t,t_0)$ and $\tilde{\lambda}^{\eta,\etap}(t,t_0)$ from solving the GEVP on $\mathcal{C}(t)$ and $\tilde{\mathcal{C}}(t)$, respectively. In particular, for the $\etap$ state which contains a much larger singlet component than the $\eta$, there is a significant shift below zero in $\lambda^{\etap}(t,t_0)$ at large times that is absent in $\tilde{\lambda}^{\etap}(t,t_0)$. Moreover, the point errors for $\tilde{\lambda}^{\etap}(t,t_0)$ are reduced by an order of magnitude compared to $\lambda^{\etap}(t,t_0)$. \par

\begin{figure}[t]
 \centering
 \includegraphics[totalheight=0.226\textheight]{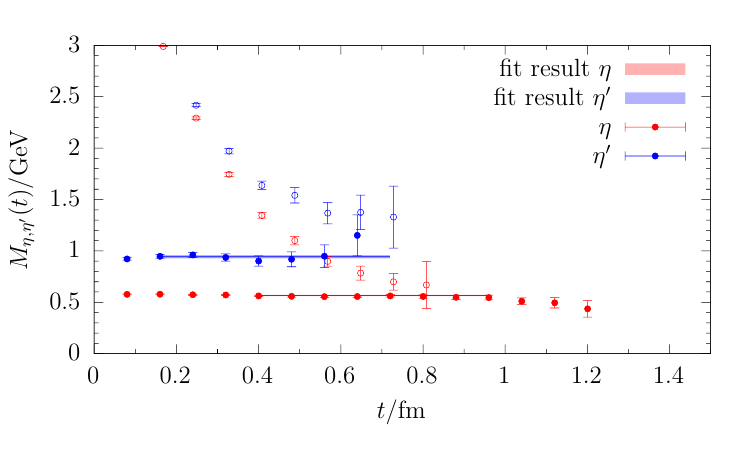}
 \includegraphics[totalheight=0.226\textheight]{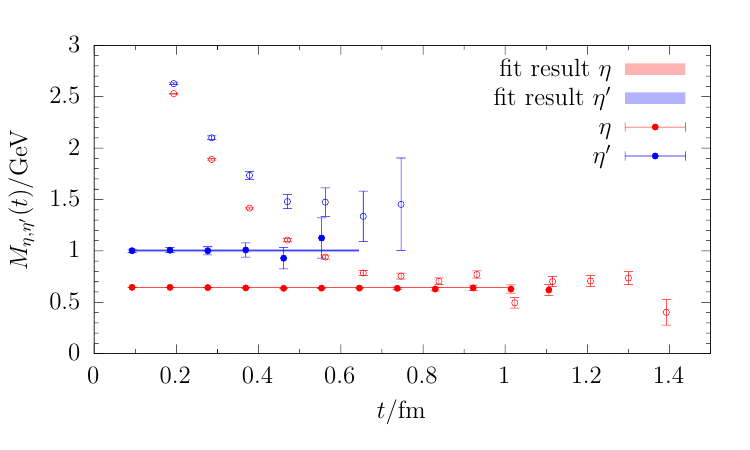}
 \caption{Comparison of the effective masses for the $\eta$ and $\etap$ from the standard approach (open symbols) and with excited states removed in the quark-connected contributions (filled symbols). Results for $M_{\eta,\etap}$ are obtained from fits to the correponding principal correlators and are shown by the solid lines and the (barely visible) error bands. The respective fit ranges are indicated by the start and endpoint of the lines. Left panel: cB211.072.64, right panel: cA211.53.24.}
 \label{fig:ESRM_eff_mass_comparison}
\end{figure}

In Fig.~\ref{fig:ESRM_eff_mass_comparison} effective masses $\Meta(t)$ and $\Metap(t)$ are shown for two ensembles. A comparison of the results from solving the $2\times2$ GEVP for correlation function matrices $\tilde{\mathcal{C}}(t)$ with and without excited states removed. The efficacy of removing the excited-state contamination in the quark-connected contributions is clearly demonstrated as a plateau is approached at very early values of $t/a$. The lattice results for the masses $a\Meta$ and $a\Metap$ have been collected in Table~\ref{tab:M_eta_and_M_etap}. \par

\subsection{Valence strange-quark mass matching} \label{subsec:mu_s_matching}
For observables that involve valence strange-quark contributions the individual results at fixed values $\mu_s^{\mathrm{val},i}$ first need to be interpolated to a suitable target value before the physical extrapolation can be carried out. In order to obtain the required target values for the strange-quark mass at every value of $\beta$, we have implemented three different choices for the matching condition labelled by M1, M2 and M3, i.e.,
\begin{align}
 \text{M1:} \ & \ M_\Omega[a^2,M_\pi^\phys,\mu_s^\text{M1}] = M_\Omega^\phys \,, \label{eq:M1} \\ 
 \text{M2:} \ & \ M_K[a^2,M_\pi^\phys,\mu_s^\text{M2}] = M_K^\phys  \,, \label{eq:M2} \\
 \text{M3:} \ & \ M_{\eta_s}[a^2,M_\pi^\phys,\mu_s^\text{M3}] = M_{\eta_s}^\phys  \,, \label{eq:M3} 
\end{align}
where $M_\Omega$ denotes the mass of the $\Omega^-$ baryon. For the physical masses of the pion and kaon we use the values in the isospin-symmetric limit $M_\pi^\phys = 134.8\mev$ and $M_K^\phys=494.2\mev$ from Ref.~\cite{Aoki:2016frl}. The physical value for the mass of the $\Omega^-$ baryon $M_\Omega^\phys = 1672.45(29)\mev$ is taken from Ref.~\cite{ParticleDataGroup:2018ovx}, whereas the physical mass of the artificial $\eta_s$ meson $M_{\eta_s}^\phys=689.89\mev$ has been determined in Ref.~\cite{Borsanyi:2020mff} to high statistical precision. For the three finest lattice spacings the target values $\mu_s^\text{M1,\ldots,M3}$ are determined directly from a linear fit of the measured values $M_X[M_\pi^\phys, \mu_s^{\mathrm{val},i}]$, $X=\Omega,K,\eta_s$ as a function of $\mu_s^\mathrm{val}$ using the lattice data obtained on the physical quark-mass ensembles (i.e., cB211.072.64, cC211.06.80 and cD211.054.96) at the three values $\mu_s^{\mathrm{val},i}$. For the kaon and the $\eta_s$ these calculations have been carried out for the same values of $\mu_s^{\mathrm{val},i}$ in Table~\ref{tab:beta_dependent_parameters} and matching gauge statistics with the other parts of the analysis.
However, for technical reasons the measurements of $aM_\Omega$ were performed on a different set of $\mu_s^{\mathrm{val},i}$ values, which are compiled in Table~\ref{tab:M_Omega} and only on a subset of ensembles as well as a smaller number of gauge configurations. \par

Since we do not have an ensemble with physical quark masses at the coarsest lattice spacing ($\beta=1.726$), we first carry out a linear fit of $M_X[M_\pi^2, \mu_s^{\mathrm{val},i}]$ in $M_\pi^2$ to obtain the values $M_X[M_\pi^\phys, \mu_s^{\mathrm{val},i}]$ at physical light-quark mass at each of the three values of $\mu_s^{\mathrm{val},i}$ for this $\beta$-value. In a second step, we apply the same procedure to these values $M_X[M_\pi^\phys, \mu_s^{\mathrm{val},i}]$ as for the other values of the lattice spacing to determine the target values $\mu_s^\text{M1,\ldots,M3}$. In order to keep the uncertainty introduced by this additional step as small as possible, we restrict the input data for the extrapolation in the light-quark mass to the three most chiral ensembles, i.e., cA211.12.48, cA211.30.32 and cA211.40.24, where cA211.15.48 had to be excluded due to a lack of measurements for $M_\Omega$, cf.~Table~\ref{tab:M_Omega}. The resulting target values $\mu_s^\text{M1,\ldots,M3}$ have been collected in Table~\ref{tab:beta_dependent_parameters}. \par

Finally, the results for any observable $O$ are corrected to the respective $\mu_s$ target values on each gauge ensemble. Again, this is achieved by fitting the results for $O$ that have been obtained at the three values $\mu_s^{\mathrm{val},i}$ as a linear function of $\mu_s^\mathrm{val}$ and subsequently using the fit parameters to compute its target value at $\mu_s^\text{M1,\ldots,M3}$. This entire procedure is carried out on every bootstrap sample to propagate all errors towards the final results. \par

\begin{figure}[t]
 \centering
 \includegraphics[totalheight=0.225\textheight]{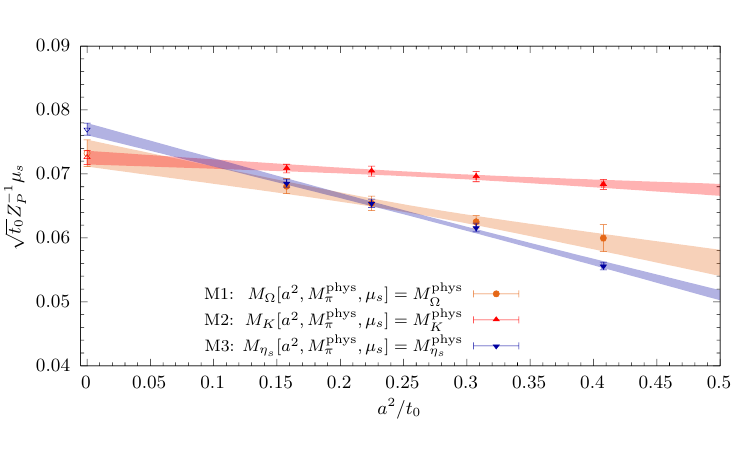}
 \includegraphics[totalheight=0.225\textheight]{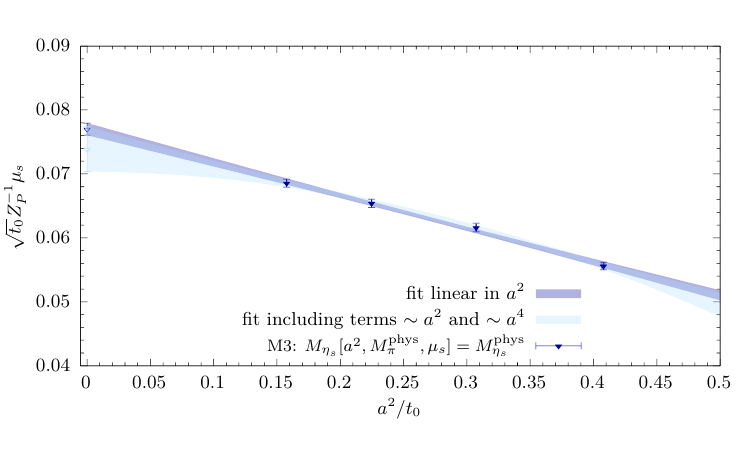}
 \caption{Target values for the valence strange-quark mass as a function of $a^2/t_0$. Left panel: Results for the three matching conditions fitted as a function of $a^2/t_0$. Right panel: Comparison of a fit linear in $a^2/t_0$ and a fit including a term of $\mathcal{O}(a^4)$ for the data obtained from matching condition M3.}
 \label{fig:mu_s_target_values}
\end{figure}

In the left panel of Fig.~\ref{fig:mu_s_target_values} renormalized results for the target strange-quark mass values are shown as a function of $a^2/t_0$ together with a linear fit for each of the three matching conditions. We find that results for M1 and M2 are in excellent agreement although they approach the continuum limit with a very different slope in $a^2/t_0$. For M3 we observe a $\sim 2-3 \sigma$ tension with the results from methods M1 and M2. This can likely be attributed to higher-order effects in the lattice spacing, which is demonstrated in the right panel of Fig.~\ref{fig:mu_s_target_values} comparing the linear fit for M3 to a fit including a quadratic term $\sim a^4$. The latter results in a lower continuum value for $\sqrt{t_0} Z_P^{-1}\mu_s^\text{M3}$ in good agreement with the other two matching conditions, albeit within increased statistical errors. \par

However, it is important to note that a steep continuum extrapolation for $\mu_s$ itself does not necessarily imply the same for other observables. Still, different choices of the matching condition generally result in different continuum extrapolations. This is why a physical extrapolation is carried out for the results from all three matching conditions instead of making an explicit choice. Instead, we prefer to combine the individual results in a model average to account for systematic effects due the choice of the matching condition. \par

\section{Physical extrapolations}
\label{sec:physical_extrapolations}

After implementing any given matching condition for $\mu_s^\mathrm{val}$ by correcting the lattice data to the corresponding target value of $\mu_s^\mathrm{val}$ on every ensemble, the physical extrapolations remain to be carried out. To this end, we carry out global fits for every observable $O$, that are variations of a simple fit ansatz inspired by chiral perturbation theory
\begin{equation}
 O^n(M_\pi, a) = \chiral{O}_\text{SU(2)}^n + A_O M_\pi^2 + B_O a^2 \,,
 \label{eq:CCF}
\end{equation}
where $n=2$ for $O=\Meta,\Metap$ and $n=1$ otherwise. $\chiral{O}_\text{SU(2)}$ denotes the value of the observable in the SU(2) chiral limit, and $M_\pi^2 \sim m_l$ is used as a proxy for the light-quark mass. Since the sea strange-quark mass has been tuned sufficiently well to its physical value for the purposes of the present study, cf. Ref.~\cite{ExtendedTwistedMassCollaborationETMC:2024xdf}, the model does not include a dependence on the strange-quark mass. \par

The above expression only accounts for the chiral and continuum extrapolations but not the finite-volume effects. Depending on the observable they are treated in two different ways. For $M_\pi$ and $\Meta$ we explicitly correct the lattice data by applying the analytic finite-volume corrections obtained in Ref.~\cite{Colangelo:2005gd} from the resummed L\"uscher formulae. The relation between the infinite-volume result $M_P$ for $P=\pi,\eta$ and its finite-volume counterpart $M_P(L)$ is given by 
\begin{equation}
 M_P=\frac{M_P(L)}{1+R_{M_P}} \,,
 \label{eq:infinite_volume_observables}
\end{equation}
where the relevant expressions for $R_{M_P}$ read
\begin{align}
 R_{M_\pi}  &= \frac{1}{4} \xi_\pi \tilde{g}_1(\lambda_\pi) - \frac{1}{12} \xi_\eta \tilde{g}_1(\lambda_\eta) \,, \label{eq:R_M_pi} \\
 R_{\Meta} &= \frac{1}{2} \xi_K \tilde{g}_1(\lambda_K) - \frac{1}{3} \xi_\eta \tilde{g}_1(\lambda_\eta) - \frac{M_\pi^2}{\Meta^2} \l( \frac{1}{4} \xi_\pi \tilde{g}_1(\lambda_\pi) - \frac{1}{6} \xi_K \tilde{g}_1(\lambda_K) - \frac{1}{12} \xi_\eta \tilde{g}_1(\lambda_\eta) \r) \,, \label{eq:R_M_eta} 
\end{align}
with shorthands $\xi_P=\frac{M_P^2}{(4\pi f_\pi)^2}$, $\lambda_P=M_P L$, and
\begin{equation}
 \tilde{g}_1(x)=\sum_{n=1}^{\infty} \frac{4m(n)}{\sqrt{n} x} K_1\l(\sqrt{n}x\r) \,,
\end{equation}
where $K_1$ denotes the Bessel function of the second kind and $m(n)$ are the multiplicities tabulated in Ref.~\cite{Colangelo:2003hf}. We do not need corrections to $M_K$ here, as it does not enter any of the global fits. Besides, we remark that there are corresponding expressions for $f_\pi$, $f_K$ and $f_\eta$ given in Ref.~\cite{Colangelo:2005gd}. However, the definition of $f_\eta$ assumes that the $\eta$ is a pure octet state defined in the absence of mixing, hence this result cannot be used for the decay-constant parameters $f_l$ and $f_s$ in the quark-flavor basis. Besides, $f_\pi$ and $f_K$ are only required for the ratios $f_l/f_\pi$ and $f_s/f_K$ which we consider to cancel some of the finite-volume effects and / or scaling artifacts. Therefore, we choose not to correct our data for $f_\pi$ and $f_K$ that enter these ratios. \par

The finite-volume corrections in Eqs.~(\ref{eq:infinite_volume_observables})~and~(\ref{eq:R_M_eta}) are applied to the lattice data on every ensemble after the valence strange-quark matching procedure described in Subsec.~\ref{subsec:mu_s_matching} has been carried out. The corrections are found to be rather small, even for $M_\pi$ they are typically at most a few permille on our ensembles. The only exception is the cB211.25.24 ensemble for which the correction reaches the percent level. On the other hand, for ensembles with large volumes the corrections become a sub-permille effect. For $\Meta$ they are entirely negligible within the present statistical precision as the relative correction is $<10^{-3}$ even on the cB211.25.24 ensemble. \par

For any fitted observable other than $\Meta$ we vary the fit model in Eq.~(\ref{eq:CCF}) by complementing it with an optional term 
\begin{equation}
 C_0 \frac{M_\pi^2}{\sqrt{M_\pi L}} e^{-M_\pi L} \,,
 \label{eq:FV_term}
\end{equation}
with a free, observable-dependent fit parameter $C_O$ to test for finite-volume effects. We still allow models without such a term to enter the final model averages because in many cases finite-volume effects cannot be resolved within the statistical precision of our data. \par

In order to corroborate the chiral extrapolation and to include possible higher-order effects into our systematic error estimates, we consider further variations by adding a term
\begin{equation}
 X_O \in \l\{D_O M_\pi^2\log{M_\pi}, \ E_O M_\pi^4 , \  F_O  M_\pi^4 \log{M_\pi} \r\} \,,
 \label{eq:higher_order_terms}
\end{equation}
to any of the previously defined fit models. Again, the coefficients $D_O$, $E_O$ and $F_O$ are treated as free parameters of the fit. Moreover, we apply three different data cuts~$\in\{M_\pi<270\mev, a\leq 0.08\fm, M_\pi L<3.5\}$ for every fit model to test for residual corrections to any of the extrapolations in $M_\pi^2$, $a^2$ and $L$. \par

\subsection{Masses} \label{subsec:CCF_results_masses}
\begin{figure}[t]
 \centering
 \includegraphics[totalheight=0.226\textheight]{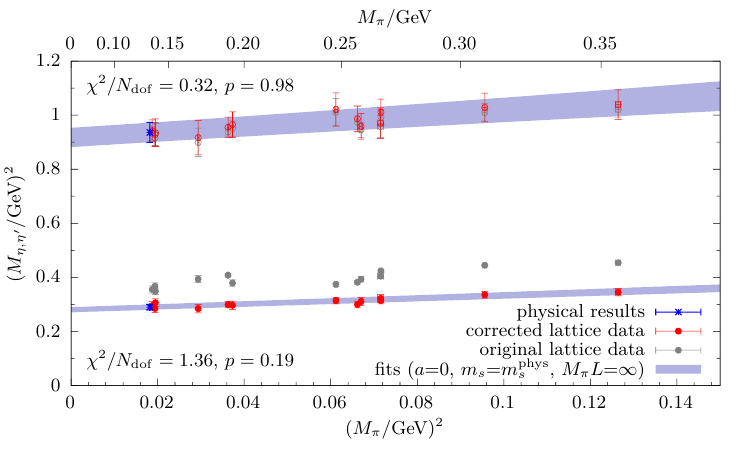}
 \includegraphics[totalheight=0.226\textheight]{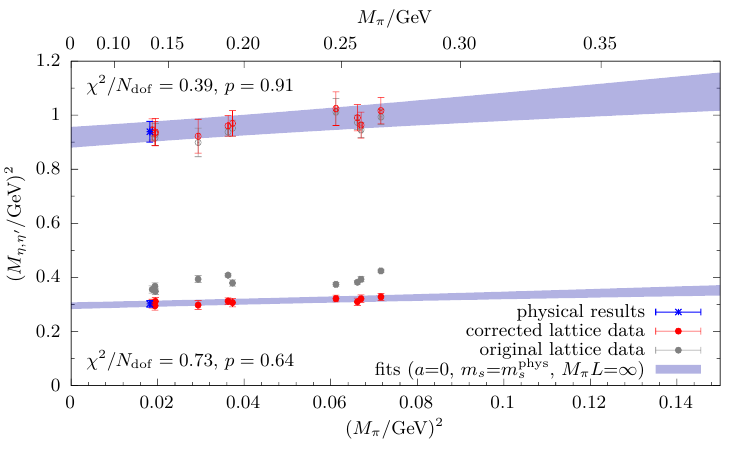} \\
 \includegraphics[totalheight=0.226\textheight]{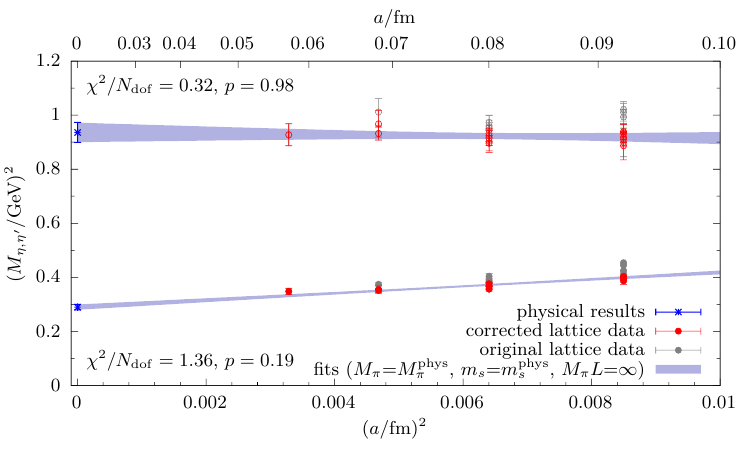}
 \includegraphics[totalheight=0.226\textheight]{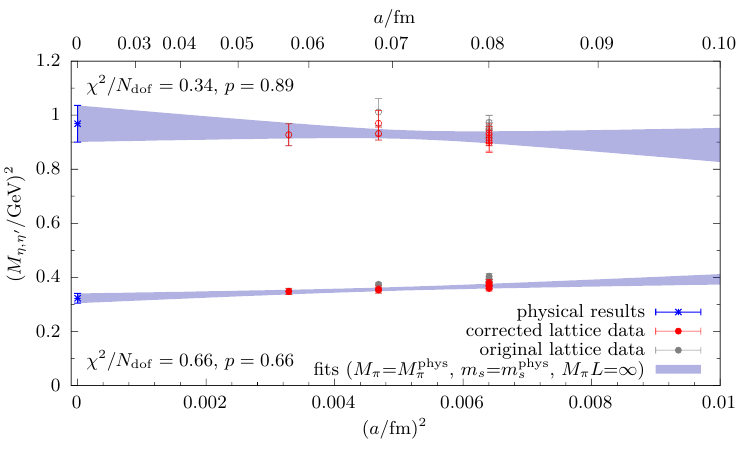}
 \caption{Physical extrapolations for $\Meta$ and $\Metap$ obtained from the fit model in Eq.~(\ref{eq:CCF}). Top row: chiral extrapolation in the light-quark mass for the full data set (left panel) and with a cut in $M_\pi\lesssim 270\mev$ (right panel). Bottom row: continuum extrapolation for the full data set (left panel) and with a cut in $a\lesssim 0.08\fm$. The ``original'' data (gray symbols) correspond to the $\mu_s^\text{M1}$ target values in Table~\ref{tab:beta_dependent_parameters}. The red data have been corrected for the physical extrapolation in all variables besides the one on the horizontal axis using the parameters from the fit. Therefore, the respective point errors are highly correlated. Errors are statistical only.}
 \label{fig:CCF_M_eta_and_M_etap}
\end{figure}

\begin{figure}[t]
 \centering
 \includegraphics[totalheight=0.226\textheight]{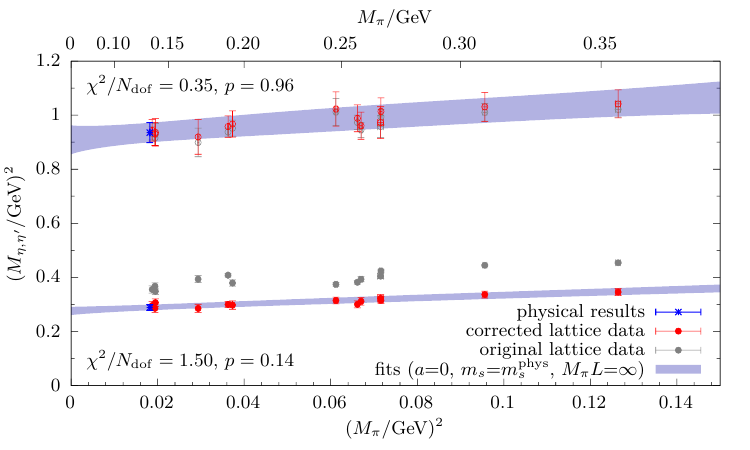}
 \includegraphics[trim=0 0.925em 0 -0.925em,totalheight=0.226\textheight]{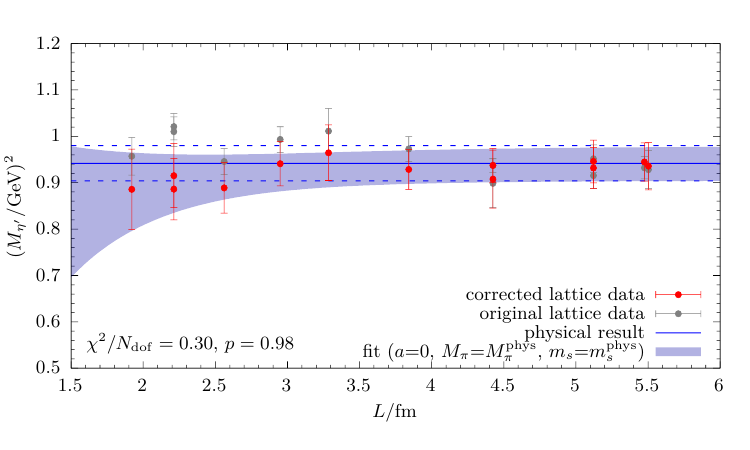}
 \caption{Two examples for variations of the model in Eq.~(\ref{eq:CCF}). Left panel: Chiral extrapolation of $\Meta^2$ and $\Metap^2$ as obtained from a fit with an additional term $\sim \Mpi^4$. Right panel: Infinite-volume extrapolation of $\Metap^2$ from a fit including the term in Eq.~(\ref{eq:FV_term}). Both fits are carried out on the full set of data from matching condition M2.}
 \label{fig:CCF_m_l_sqr_and_FV_term}
\end{figure}

Figure \ref{fig:CCF_M_eta_and_M_etap} shows results for the physical extrapolations of $\Meta$ and $\Metap$ from fitting the model in Eq.~(\ref{eq:CCF}) to lattice data obtained from the first matching condition for $\mu_s^\mathrm{val}$ in Eq.~(\ref{eq:M1}). The chiral extrapolation in the light-quark mass at $a=0$ and physical $m_s$ is found to be mild for both masses, which is in agreement with other studies \cite{Ottnad:2012fv,Michael:2013gka,Ottnad:2017bjt,Bali:2021qem}. Fitting the full set of data for $\Meta$ generally yields $p$-values $\lesssim 0.2$, depending on the choice of the matching condition for the valence strange-quark mass. The fit quality for the $\eta$ is improved by a data cut in the pion mass, i.e., $M_\pi<270\mev$, whereas it remains unaffected for the $\etap$. This is demonstrated in the upper right panel of Fig.~\ref{fig:CCF_M_eta_and_M_etap} and might be hinting at residual effects due to higher-order terms contributing in the chiral expansion in case of the much more statistically precise data for the $\eta$. Still, we do not observe a systematic improvement of the fit for the $\eta$ including any of the terms in Eq.~(\ref{eq:higher_order_terms}). An example for this is given in the left panel of Fig.~\ref{fig:CCF_m_l_sqr_and_FV_term} for a fit with a term $\sim M_\pi^4$. The resulting fit curves for $\Meta^2$ and $\Metap^2$ as functions of $M_\pi^2$ exhibit barely any deviation from a linear slope in $\Mpi^2$ and the contribution of such an additional term in the fit remains negligible. \par

\begin{figure}[t]
 \centering
 \includegraphics[totalheight=0.226\textheight]{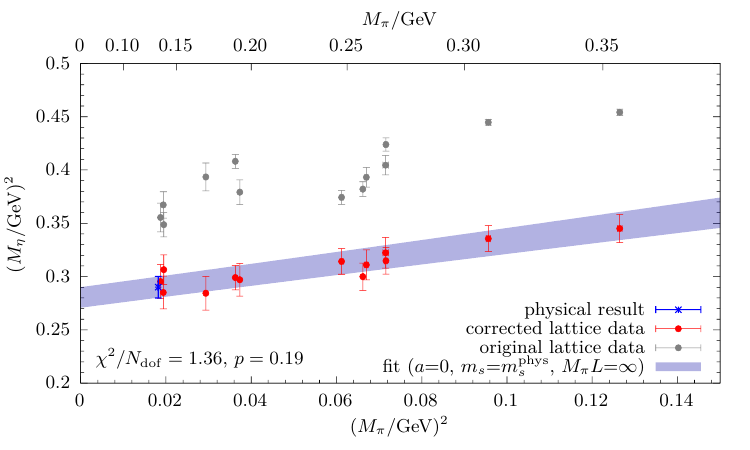}
 \includegraphics[totalheight=0.226\textheight]{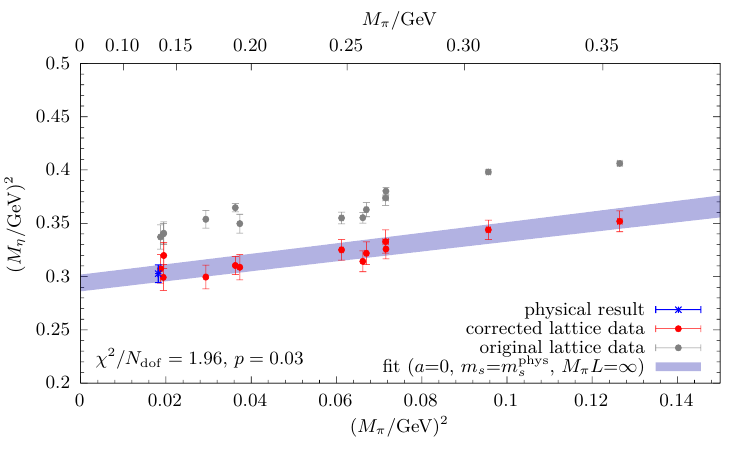}
 \caption{Comparison of the chiral extrapolation using input data from two choices of the matching condition, i.e., M2 and M3. Results have been obtained from fitting the expression in Eq.~(\ref{eq:CCF}) without modifications to the full set of data.}
 \label{fig:CCF_M_eta_M2_vs_M3}
\end{figure}

The continuum extrapolations for $\Meta^2$ and $\Metap^2$ are shown in the lower left panel of Fig.~\ref{fig:CCF_M_eta_and_M_etap} and originate from the same fit as the chiral extrapolation in the panel above it. For the $\etap$ the dependence on $a^2$ is flat and compatible with a constant extrapolation. The situation is very different for $\Meta$, which receives sizable corrections of up to $\sim 30\%$ at the coarsest lattice spacing towards the physical point. Applying a data cut in the lattice spacing improves the fit quality for the $\eta$ but not the $\etap$, similar to what has been found for a cut in the pion mass. However, within its inflated statistical errors the extrapolation remains fairly stable with compatible slope and physical result. The rather large scaling artifacts seen in $\Meta$ are related to its strong dependence on the valence (and sea) strange-quark mass. They are significantly affected by the choice of the matching condition for the target $\mu_s$ value, which has already been observed in Ref.~\cite{Ottnad:2015hva}. An example for this is shown in Fig.~\ref{fig:CCF_M_eta_M2_vs_M3} comparing results from fitting the same model to data obtained from matching methods M2 and M3. Nevertheless, all three matching methods do lead to similar chiral extrapolation curves and compatible physical results. This gives a rather strong indication that any residual, systematic effects due to the choice of the matching condition are well under control. \par 

As mentioned before, the data for $\Meta$ (and $M_\pi$) have been explicitly corrected for finite-volume effects, thus no explicit infinite-volume extrapolation is carried out for $M_\eta$. Concerning a possible volume dependence of $\Metap$, we remark that including the finite-volume term defined in Eq.~(\ref{eq:FV_term}) in the fits does not reveal any such effect. A corresponding example is given in the right panel of Fig.~\ref{fig:CCF_m_l_sqr_and_FV_term}, where the extrapolation in $L$ remains indeed entirely flat. \par

\subsection{Mixing parameters} \label{subsec:CCF_results_mixing}
Just like the masses $\Meta$ and $\Metap$, the mixing angle $\phi$ as well as the decay-constant parameters $f_l$, $f_s$ defined through Eq.~(\ref{eq:PS_mixing}) generally depend on the valence strange-quark mass $\mu_s^\mathrm{val}$; cf. Tables~\ref{tab:phi}~and~\ref{tab:f_l_and_f_s} in the appendix. Therefore, the lattice data must again be corrected to the target $\mu_s$ values derived from the matching conditions in Eqs.~(\ref{eq:M1})-(\ref{eq:M3}) before the physical extrapolation can be carried out. However, for the mixing parameters there is a subtle ambiguity concerning the implementation of the corresponding procedure. This is because in principle one could implement any matching condition already at the level of the matrix elements $h^{\eta,\etap}_{l,s}$ as extracted from the correlation functions, i.e., before evaluating the mixing parameters from Eq.~(\ref{eq:PS_mixing}). Still, with respect to preserving correlations we find it somewhat beneficial to postpone the matching until the actual observables have been computed. In particular, the mixing angle is obtained from a double ratio of matrix elements
\begin{equation}
 \phi = \arctan \sqrt{-\frac{h^{\etap}_l h^\eta_s}{h^\eta_l h^{\etap}_s}} \,,
 \label{eq:phi_double_ratio}
\end{equation}
which is sensitive to cancellations of the highly correlated fluctuations between the individual matrix elements. \par

\begin{figure}[t]
 \centering
  \includegraphics[totalheight=0.226\textheight]{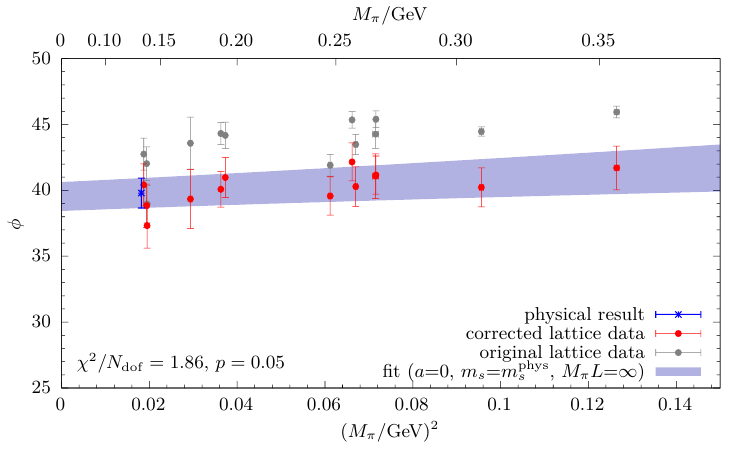}
  \includegraphics[totalheight=0.226\textheight]{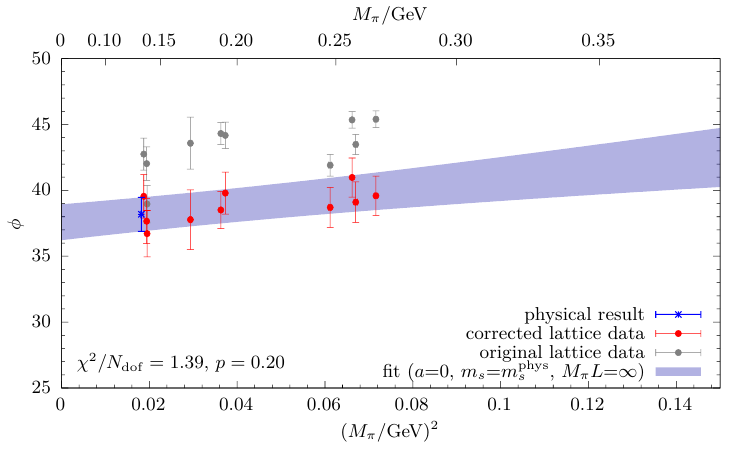} \\
  \includegraphics[totalheight=0.226\textheight]{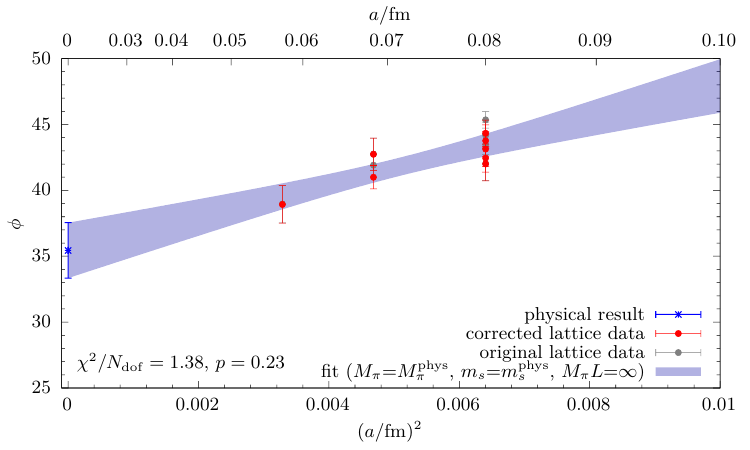}
  \includegraphics[trim=0 0.925em 0 -0.925em,totalheight=0.226\textheight]{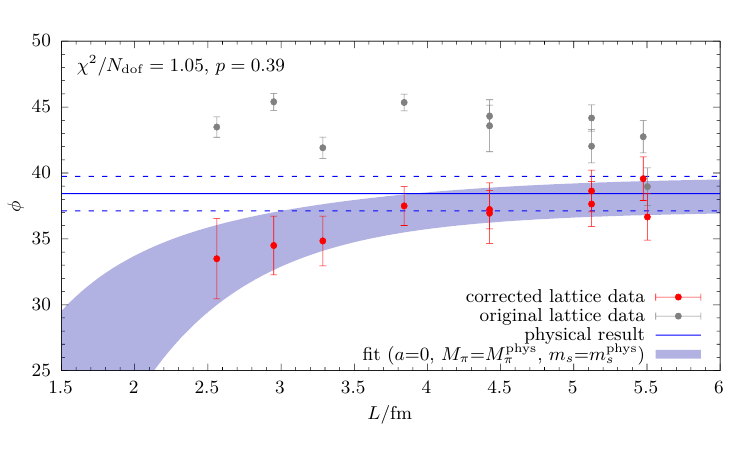}
  \caption{Examples for the physical extrapolation of the mixing angle $\phi$ based on input data from matching method M3 in Eq.~(\ref{eq:M3}). The chiral extrapolation from a fit to the model in Eq.~(\ref{eq:CCF}) is shown for the full set of data (upper left panel) and for a pion mass cut of $M_\pi\leq270\mev$ (upper right panel). The continuum extrapolation in the lower left panel is obtained from a fit to the same model but with a data cut in the lattice spacing $a\leq0.08\fm$. For the infinite-volume extrapolation in the lower right panel the term in Eq.~(\ref{eq:FV_term}) has been added to the model, while the input data set remains them same as in the upper right panel.}
  \label{fig:CCF_phi}
\end{figure}

In Fig.~\ref{fig:CCF_phi} exemplary results are shown for the physical extrapolation of the mixing angle $\phi$. We find that fits of Eq.~(\ref{eq:CCF}) to the full set of data typically yield small $p$-values, i.e., $p\lesssim0.5$. Data cuts in the pion mass or the lattice spacing generally improve the fit quality, as can be seen in the upper right and lower left panel of Fig.~\ref{fig:CCF_phi}. The situation is significantly improved further by including the term in Eq.~(\ref{eq:FV_term}) in the fit, indicating non-negligible finite-volume effects in $\phi$. In fact, the corresponding fit parameter $C_\phi$ is found to be negative within $~1$ to $~2\sigma$ for every single tested combination of fit-model variations, data cuts and $\mu_s$-matching conditions. On the other hand, none of the terms in Eq.~(\ref{eq:higher_order_terms}) that are of higher order in the chiral expansion improve the fit on their own. They only affect the fit in certain combinations with the finite-volume term in Eq.~(\ref{eq:FV_term}) and data cuts. Such combinations tend to overfit the available data due to the small number of degrees of freedom remaining. Considering the flatness of the chiral extrapolation and its linear behavior in $M_\pi^2$, we conclude that higher-order corrections in the light-quark mass are likely negligible. Finally, concerning the continuum extrapolation, we remark that corrections can be as large as $20\%$, depending on the lattice spacing and the choice of the matching condition for the valence strange-quark mass.\par

\begin{figure}[t]
 \centering
  \includegraphics[totalheight=0.226\textheight]{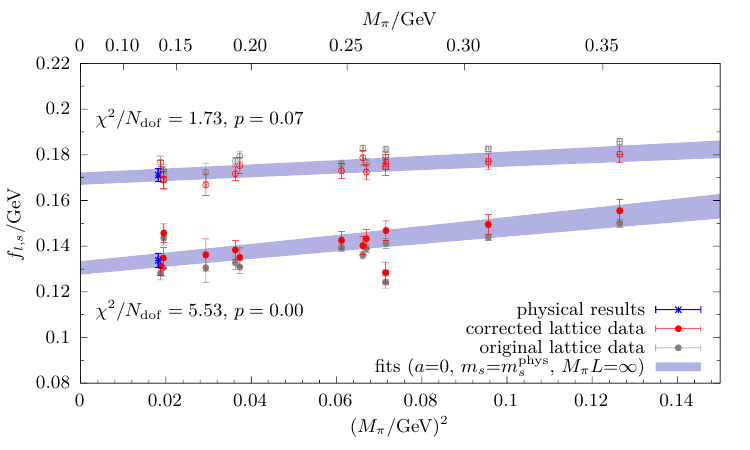}
  \includegraphics[totalheight=0.226\textheight]{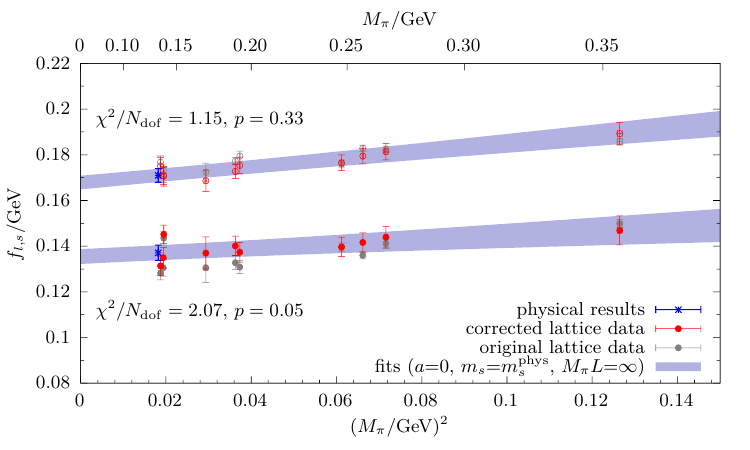} \\
  \includegraphics[totalheight=0.226\textheight]{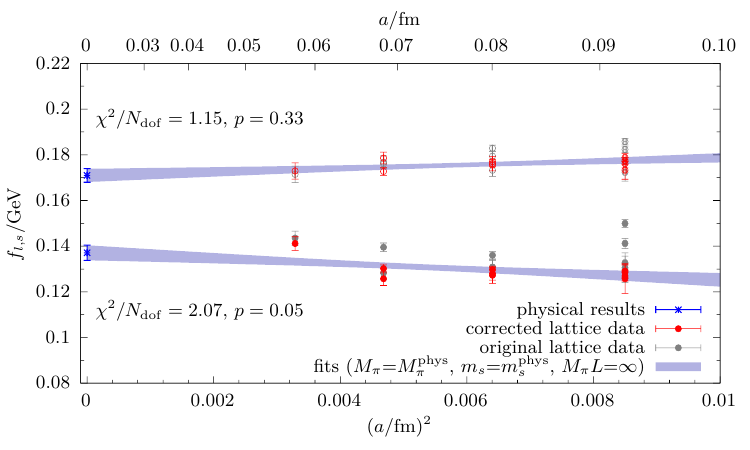}
  \includegraphics[trim=0 0.925em 0 -0.925em,totalheight=0.226\textheight]{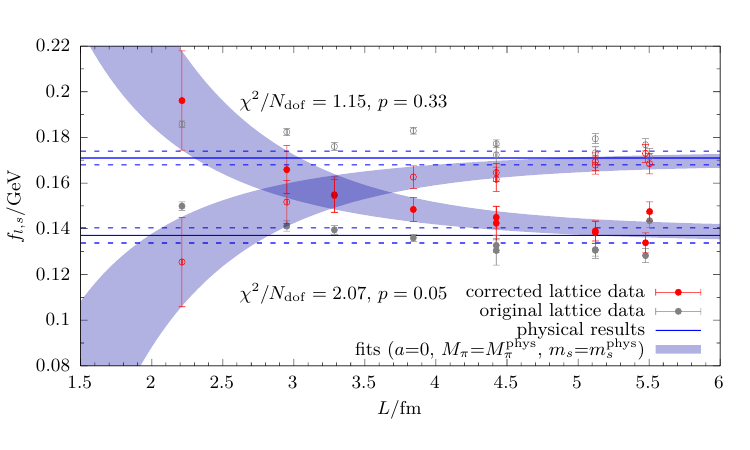}
  \caption{Examples for the physical extrapolation of the decay-constant parameters $f_l$ and $f_s$ based on input data from $\mu_s^\mathrm{val}$-matching method M1 in Eq.~(\ref{eq:M1}). The chiral extrapolation in the upper left panel results from a fit of Eq.~(\ref{eq:CCF}) to the full set of data, whereas for the remaining three panels the term in Eq.~(\ref{eq:FV_term}) has been included in the fit, and a cut of $M_\pi L<3.5$ has been applied to the input data.}
  \label{fig:CCF_f_l_and_f_s}
\end{figure}

For the decay-constant parameters the most basic model in Eq.~(\ref{eq:CCF}) results in very different fit qualities for $f_l$ and $f_s$, as can be seen in the upper left panel of Fig.~\ref{fig:CCF_f_l_and_f_s}. Unlike for the masses and the mixing angle, we find that data cuts in $\Mpi$ or $a$ lead to an improvement neither for $f_l$ nor $f_s$. On the other hand, a cut in $\Mpi L < 3.5$ and inclusion of the finite volume term in Eq.~(\ref{eq:FV_term}) reduces the correlated $\chi^2/N_\mathrm{dof}$ for both of them, although the effect is much more drastic for $f_l$ in the upper right panel of Fig.~\ref{fig:CCF_f_l_and_f_s}. While corrections due to the continuum extrapolations in the lower left panel of Fig.~\ref{fig:CCF_f_l_and_f_s} remain at the few-percent-level, both observables are affected by significant finite volume effects below $L\lesssim 4.5\fm$ as shown in the lower right panel. The corresponding fit parameters are different from zero by several standard deviations and have opposite sign. More precisely, for $f_s$ we find $C_{f_s}<0$ for every single fit that includes the term in Eq.~(\ref{eq:FV_term}), whereas $C_{f_l}>0$ for every fit with $p\geq0.01$. In general, it is not possible to obtain a reasonable description of our $f_l$ data by any of the fit models if the cB211.25.24 ensemble is included in the fit. The reason for this that for all three choices of $\mu_s^\mathrm{val}$ the value for $f_l$ is $\sim5\sigma$ smaller than the corresponding values on cB211.25.32 and cB211.25.48; cf. Table~\ref{tab:f_l_and_f_s}. Since cB211.25.24 exhibits the smallest physical volume ($L=1.92\fm$) as well as the smallest value of $\Mpi L =2.64$ among all ensembles, this may give a hint that finite volume effects are even more severe for $f_l$ than $f_s$ to the point that they cannot be described by the naive term in Eq.~(\ref{eq:FV_term}) anymore. \par

However, even for fits that include the term in Eq.~(\ref{eq:FV_term}) and omit the data for cB211.25.24, the fit quality for $f_l$ is always worse than for $f_s$. While we cannot exclude that this may be partially caused by residual finite-volume effects, a closer inspection of the data reveals that it is mostly due to the cD211.054.96 ensemble. The (corrected) values for $f_l$ on this ensemble typically deviates by one to three $\sigma$ from the extrapolation curves, depending on the fit model and extrapolation variable. In Fig.~\ref{fig:CCF_f_l_and_f_s} this can be clearly seen in the continuum extrapolation in the lower left panel in which the extrapolation curve misses the data point at the smallest lattice spacing. Since such a deviation is not observed in any other observable including $f_s$, and because there is no such trend for the other two ensembles at physical quark mass, we conclude that this is most likely to be considered a statistical fluctuation. \par

\section{Model average and final results} \label{sec:AIC_final_results}
The final results for each observable are computed via a model averaging procedure that is based on a variant of the Akaike information criterion (AIC) \cite{Akaike1998,1100705}. More specifically, we employ the Bayesian AIC (BAIC) introduced in Ref.~\cite{Neil:2022joj}
\begin{equation}
 B_{m,b} = \chi^2_{m,b} + 2 N_{\mathrm{par},m} + 2 N_{\mathrm{cut},m} \,,
 \label{eq:BAIC}
\end{equation}
in the definition of the weight \cite{doi:10.1177/0049124104268644,BMW:2014pzb,Neil:2022joj} 
\begin{equation}
 w_{m,b} = \frac{e^{-B_{m,b}/2}}{\sum_{i=1}^{N_M} e^{-B_{i,b} / 2}} \,,
 \label{eq:weights}
\end{equation}
that we assign to each model $m\in\{1,...,N_M\}$ on a bootstrap sample $b \in \{1,...,N_B\}$. The BAIC itself only depends on the minimized, correlated $\chi^2$ denoted by $\chi^2_{m,b}$, the number of fit parameters $N_{\mathrm{par},m}$, and the number of data points that are removed by a possible data cut $N_{\mathrm{cut},m}$ applied in the $m$th model. Since we do not use any priors in the analysis, a corresponding term is absent in the definition of $B_{m,b}$. Following the method introduced in Ref.~\cite{Borsanyi:2020mff}, we define the cumulative distribution function (CDF)
\begin{equation}
  \mathrm{CDF}(y)= \frac{1}{N_B} \sum_{m=1}^{N_M} \sum_{b=1}^{N_B} w_{m,b} \Theta(y-O_{m,b}) \,,
  \label{eq:CDF}
\end{equation}
where $\Theta(x)$ is the Heaviside step function and $O_{m,b}$ denotes the result for any observable $O$ for the $m$th model on the $b$th bootstrap sample. The central value for each observable is given by the median obtained from the respective CDF, and the total error is defined by its quantiles that correspond to $1\sigma$ errors in case of a Gaussian distribution. We remark that we evaluate the CDF in a fully non-parametric way by using the bootstrap distributions directly, unlike the parametric procedure described in Ref.~\cite{Borsanyi:2020mff} that resorts to parametric (Gaussian) CDFs for the individual models derived from their respective central values and statistical errors. The separation of statistical and systematic errors can be carried out in an analogous way as described in Ref.~\cite{Borsanyi:2020mff} using a factor of $\lambda=2$ in the rescaling of the statistical error, which we implement at the level of the bootstrap distributions. \par

\begin{figure}[t]
 \centering
  \includegraphics[totalheight=0.209\textheight]{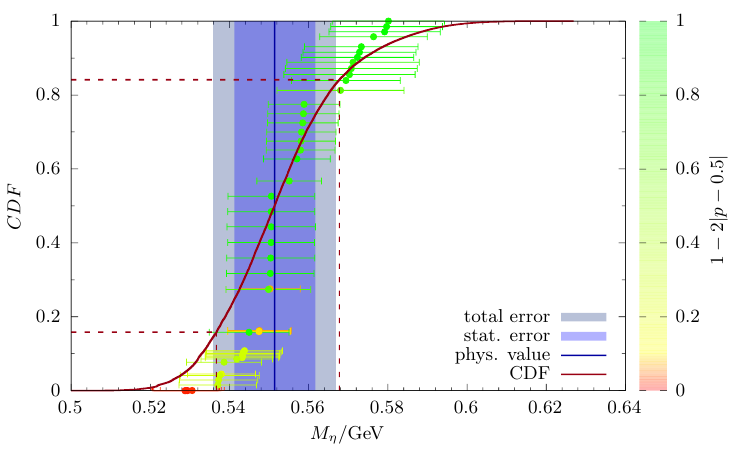}
  \includegraphics[totalheight=0.209\textheight]{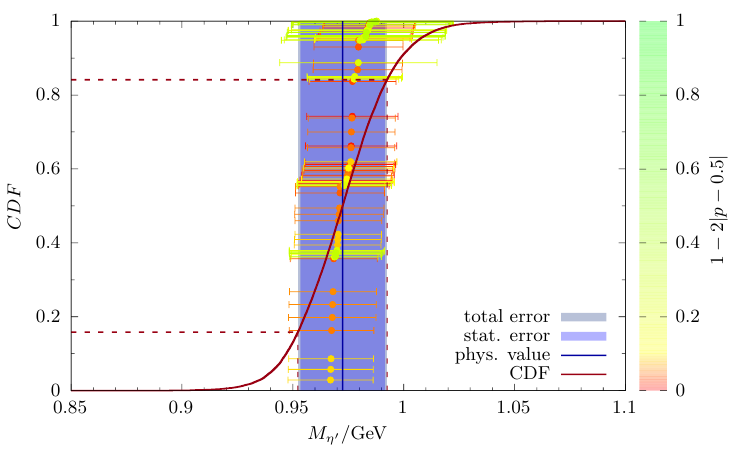}
  \caption{Cumulative distribution functions (CDFs) of the fit models for $\Meta$ (left panel) and $\Metap$ (right panel). Each data point corresponds to the central value and statistical error of one particular fit model. Their color is indicative of the respective $p$-value as defined in the plot to give some hint on the quality of individual fits. The final results from the AIC model average are indicated by the solid blue line together with its statistical and full error bands as indicated in the plots. The $1\sigma$-quantiles of the CDF are shown by dashed lines.}
  \label{fig:CDF_M_eta_M_etap}
\end{figure}

The CDFs for $\Meta$ and $\Metap$ are shown in Fig.~\ref{fig:CDF_M_eta_M_etap} and the final, physical results for the masses read
\begin{align}
 \Meta^\phys  &= 551\staterr{10}\syserr{12}\totalerr{16}\mev \,, \label{eq:M_eta_phys} \\
 \Metap^\phys &= 972\staterr{19}\syserr{06}\totalerr{20}\mev \,. \label{eq:M_etap_phys}
\end{align}
These results are in excellent agreement with experiment and the few other lattice studies that include a physical extrapolation. This includes our previous studies \cite{Michael:2013gka,Ottnad:2017bjt} on an older set of $N_f=2+1+1$ ensembles without simulations at physical quark mass ($\Mpi\gtrsim220\mev$), as well as the more recent ones by the Regensburg group in Ref.~\cite{Bali:2021qem} on $N_f=2+1$ ensembles with Wilson-clover quarks provided by the Coordinated Lattice Simulation consortium (CLS) and the one based on rooted staggered quarks in Ref.~\cite{Verplanke:2024msi}. \par 

It is worth noting that compared to our most recent previous study~\cite{Ottnad:2017bjt}, the total uncertainty in $\Metap$ could be reduced by about a factor three.
In particular, the error is no longer dominated by the statistical uncertainty, and the systematic error is much better controlled.
The errors for $\Metap$ quoted in Ref.~\cite{Bali:2021qem}, on the other hand, are significantly smaller, which we attribute to their extrapolation procedure, which is strongly constraining $\Metap$ by other quantities through a simultaneous fit. The quoted total uncertainty for $\Metap$ in Ref.~\cite{Verplanke:2024msi} is about a factor two larger than the one we quote, and thus of similar size.

For $\Meta$, there is a rather distinct spread of models in the CDF, which leads to a larger systematic contribution to the total error. This can be attributed to its stronger dependence on the valence strange-quark mass and the resulting steepness of the continuum extrapolation depending on the choice of the matching condition for the valence strange-quark mass. Consequently, the error we quote for $\Meta$ is larger than the ones from Refs.~\cite{Bali:2021qem,Verplanke:2024msi}, even though the individual lattice data entering our analysis are typically of similar or even better precision. Still, the statistical and systematic errors are of similar size for $\Meta$ which gives another indication that systematic effects due to the different matching procedures are under sufficient control. On the other hand, there is little spread in the model data for $\Metap$. This is not surprising because its physical extrapolation is very stable under any variations of the fit model and data cuts as discussed in Subsec.~\ref{subsec:CCF_results_masses}. In particular, the chiral extrapolation is mild and the extrapolation in $a^2$ and $L$ are both compatible with a constant behavior. In fact, the various fit models all share a tendency to overfit the lattice data, which is reflected by the coloring of the data points in the right panel of Fig.~\ref{fig:CDF_M_eta_M_etap}. We remark, that for $\Meta$ there are only $N_M=48$ models contributing, whereas there are twice as many for any other observable including $\Metap$. The factor two difference for $\Meta$ is caused by the absence of the term in Eq.~(\ref{eq:FV_term}) in fit model variations, because we apply explicit finite-volume corrections for $\Meta$ as discussed in Sect.~\ref{sec:physical_extrapolations}.
\par

\begin{figure}[t]
 \centering
  \includegraphics[totalheight=0.209\textheight]{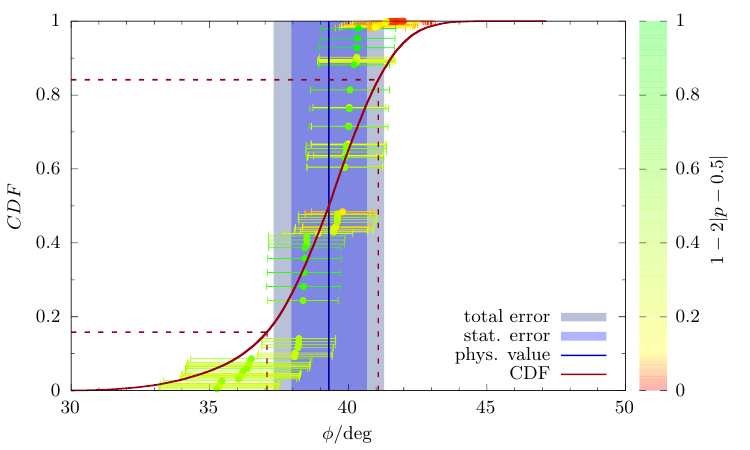}
  \caption{Same as Fig.~\ref{fig:CDF_M_eta_M_etap} but for the mixing angle $\phi$.}
  \label{fig:CDF_phi}
\end{figure}

\begin{figure}[t]
 \centering
  \includegraphics[totalheight=0.209\textheight]{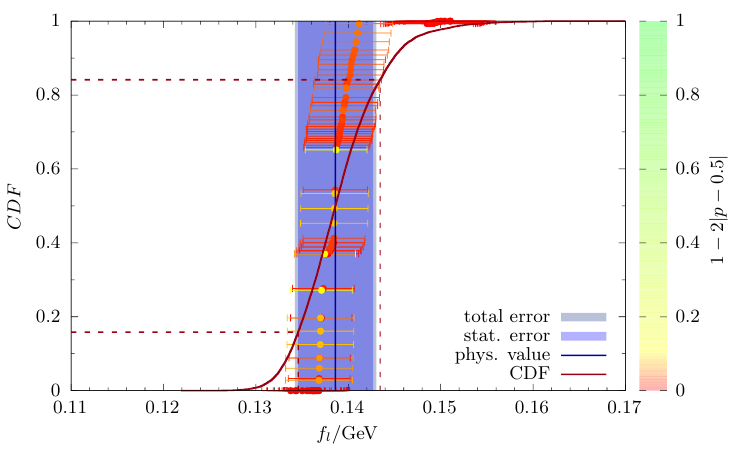}
  \includegraphics[totalheight=0.209\textheight]{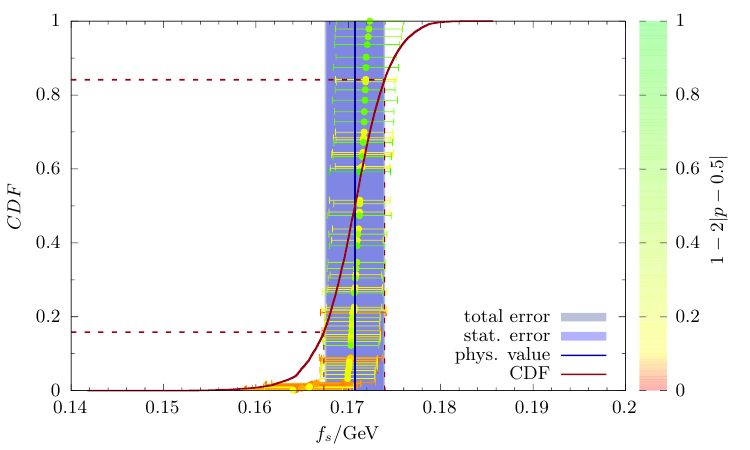}
  \caption{Same as Fig.~\ref{fig:CDF_M_eta_M_etap} but for the decay-constant parameters $f_l$ and $f_s$.}
  \label{fig:CDF_f_l_f_s}
\end{figure}

For the mixing angle $\phi$ in the FKS scheme we obtain
\begin{equation}
 \phi^\phys = {39.3\staterr{1.4}\syserr{1.5}\totalerr{2.0}}^\circ \,, \label{eq:phi_phys}
\end{equation}
in excellent agreement with our previous result $\phi={38.8\staterr{2.2}\chierr{2.4}}^\circ$ in Ref.~\cite{Ottnad:2017bjt} but with significantly reduced errors and full control over systematic effects. Specifically, the second error on the older result only accounted for the uncertainty in the chiral extrapolation. The result is also in good agreement with most phenomenological determinations; see, e.g., Refs.~\cite{Escribano:2005qq,Escribano:2013kba,Escribano:2015nra,Escribano:2015yup}. 

For the decay constant parameters $f_l$ and $f_s$ we find
\begin{align}
 f_l^\phys &= 138.6\staterr{4.1}\syserr{1.7}\totalerr{4.4}\mev \,, \label{eq:f_l_phys} \\
 f_s^\phys &= 170.7\staterr{3.1}\syserr{1.0}\totalerr{3.3}\mev \,. \label{eq:f_s_phys}
\end{align}
While $f_s^\phys$ is $\sim4\%$ smaller than the value $f_s=178\staterr{4}\chierr{1}\mev$ in Ref.~\cite{Ottnad:2017bjt} but roughly compatible within errors, we find a $\sim10\%$ larger value for $f_l^\phys$. However, this result had been affected by a particularly large uncertainty due to the chiral extrapolation, i.e., $f_l=125\staterr{5}\chierr{6}$. Furthermore, the associated error should be considered a rather crude estimate as it is just given by the difference between two fits, i.e., one to the full data set and another one with a pion-mass cut of $\Mpi\leq390\mev$. Besides, the results in Ref.~\cite{Ottnad:2017bjt} had to be computed from a physical extrapolation of the ratios $f_l/f_\pi$ and $f_s/f_K$ using the experimental values $f_\pi^\experiment=130.50\mev$ and $f_K^\experiment=155.72\mev$ in Ref.~\cite{ParticleDataGroup:2016lqr}, because a direct extrapolation of the lattice data for $f_l$ and $f_s$ had not been feasible at all. As an additional crosscheck of our analysis we have computed results for $f_l/f_\pi$ and $f_s/f_K$ in our current setup using our lattice data for $f_\pi$ and $f_K$ in Table~\ref{tab:f_pi_and_f_K}. We find 
\begin{align}
 (f_l/f_\pi)^\phys &= 1.061\staterr{29}\syserr{24}\totalerr{38} \,, \label{eq:f_l_over_f_pi_phys} \\
 (f_s/f_K)^\phys   &= 1.107\staterr{19}\syserr{12}\totalerr{23} \,. \label{eq:f_s_over_f_K_phys}
\end{align}
which yields $f_l^\phys=138.5\staterr{3.8}\syserr{3.1}\totalerr{5.0}\mev$ and $f_s^\phys=172.4\staterr{3.0}\syserr{1.9}\totalerr{3.6}\mev$, in very good agreement with the results based on the direct physical extrapolations in Eqs.~(\ref{eq:f_l_phys})~and~(\ref{eq:f_s_phys}). \par

\section{Summary and comparison with other studies}

In this paper we have presented a calculation of $\eta$ and
$\eta^\prime$ masses, the mixing angle $\phi$ in the FKS scheme, and
the corresponding decay constants $f_\ell$ and $f_s$. In summary we
find
\begin{align}
 \Meta^\phys  &= 551\totalerr{16}\mev \,,\quad
 \Metap^\phys = 972\totalerr{20}\mev \,,\quad
 \phi^\phys = {39.3\totalerr{2.0}}^\circ\,,\\
 f_l^\phys &= 138.6\totalerr{4.4}\mev \,,\quad
 f_s^\phys = 170.7\totalerr{3.3}\mev \,,
\end{align}
where we quote here only the total error, which represents the statistical
and systematic error added in quadrature.

\begin{figure}[t]
 \centering
 \includegraphics[totalheight=0.32\textheight]{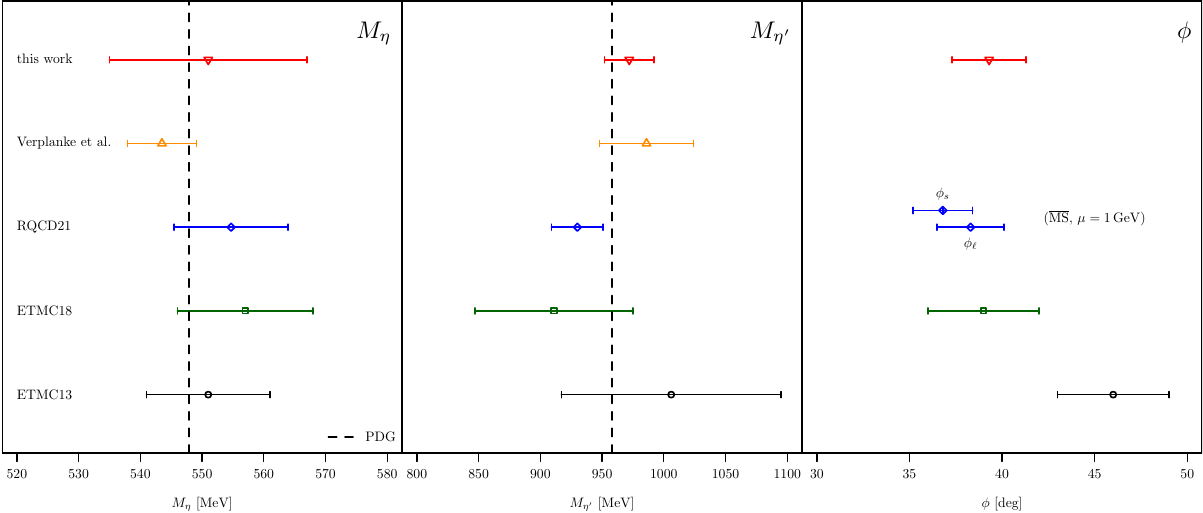}
 \caption{Comparison of the physical results for masses and mixing
   angles across different lattice calculations. We compare results
   from ETMC13~\cite{Michael:2013gka}, ETMC18~\cite{Ottnad:2017bjt},
   RQCD21~\cite{Bali:2021qem}, and Verplanke et
   al.~\cite{Verplanke:2024msi} with our results for $M_\eta$,
   $M_{\eta^\prime}$, and $\phi$. For $\phi$ we compare to $\phi_\ell$
   and $\phi_s$ from RQCD21 at $1\ \mathrm{GeV}$ in the
   $\overline{\mathrm{MS}}$-scheme. The dashed lines represent the PDG
   values for $M_\eta$ and $M_{\eta^\prime}$, respectively.}
 \label{fig:comparison}
\end{figure}

In contrast to our previous
studies with $N_f=2=1+1$ dynamical quark flavors, this calculation
includes ensembles with very close-to-physical values of the pion mass
at three values of the lattice spacing. In addition, we include
ensembles at larger-than-physical quark masses, including at a fourth
value of the lattice spacing. Therefore, an extrapolation in the light-quark
mass value was not necessary, which greatly reduces or even
removes the corresponding systematic uncertainty. Moreover, due to the 
inclusion of the clover term in the lattice action, residual lattice 
artifacts in particular stemming from flavor-breaking effects in twisted-mass
lattice QCD are significantly reduced.

We compare our results for $M_\eta$, $M_{\eta^\prime}$, and $\phi$
with other available lattice results from
Refs.~\cite{Michael:2013gka,Ottnad:2017bjt,Bali:2021qem,Verplanke:2024msi}
in Fig.~\ref{fig:comparison}. For $M_\eta$ and $M_{\eta^\prime}$, we
indicate the PDG value as the dashed vertical line. For the masses we
observe good agreement among the different lattice estimates, and also
with the respective PDG values. 

For the mixing angle $\phi$, RQCD goes beyond 
the single-angle approximation in Ref.~\cite{Bali:2021qem}, and
determines a light angle $\phi_\ell$ and a strange angle $\phi_s$,
which are additionally scale-dependent. Thus, we compare to the
results at $1\ \mathrm{GeV}$ in the $\overline{\mathrm{MS}}$-scheme
from this reference, which should be theoretically closest to the
single-angle approximation. Reassuringly, $\phi_\ell$ and $\phi_s$
agree very well with each other within errors. Moreover, there is good
agreement with our ETMC18~\cite{Ottnad:2017bjt} determination and this
work. The result for $\phi$ from ETMC13~\cite{Michael:2013gka} is
slightly larger but still well compatible within the combined errors.

The new results presented here for the mixing angle and the decay
constants are still based on pseudoscalar matrix elements. In the future it
would be interesting to estimate also the corresponding axial-vector
matrix elements, which is surprisingly difficult in Wilson twisted-mass
lattice QCD. Finally, in principle the $\etap$ should be treated as a
resonance. Calculations of this kind become doable, as can be seen for
instance from this recent study of the $\omega$ meson~\cite{Yan:2024gwp}.

\section*{Acknowledgments}
We thank all members of ETMC for the most enjoyable collaboration.
We gratefully acknowledge computing time granted by: The Gauss Centre for Supercomputing e.V. (www.gauss-centre.eu) on the GCS Supercomputers JUWELS~\cite{JUWELS,BOOSTER} at J\"ulich Supercomputing Centre as well as SuperMUC and SuperMUC-NG at Leibniz Supercomputing Centre.
We further gratefully acknowledge the computing time granted by the John von Neumann Institute for Computing (NIC) on the supercomputer JURECA~\cite{JURECA} at J\"ulich Supercomputing Centre.
We gratefully acknowledge PRACE for awarding access to HAWK at HRLS within the project with ID \emph{Acid 4886} and to Marconi at CINECA.
We further gratefully acknowledge the Swiss National Supercomputing Centre (CSCS) for awarding access to Piz Daint and to the EuroHPC JU for awarding access to the Luxembourg national supercomputer Meluxina.
We gratefully acknowledge the Texas Advanced Computing Center (TACC) at the University of Texas at Austin for proving HPC resources on Frontera (Project ID PHY21001) and the University of Bonn for granting access to the HPC cluster Bonna.
The work on this project was supported by the Deutsche Forschungsgemeinschaft (DFG, German Research Foundation) as part of the CRC 110 and the CRC 1639 NuMeriQS – project no. 511713970, and by the Swiss National Science Foundation (SNSF) through project No.~200020\_208222.
K.O. acknowledges support by the DFG through project HI~2048/1-3 (project No.~399400745). 
S.B., J.F. and F.P. received financial support from the Inno4scale project, which received funding from the European High-Performance Computing Joint Undertaking (JU) under Grant Agreement No. 101118139.
J.F.~acknowledges financial support by the Next Generation Triggers project~\cite{NextGenTriggers}.
The open source software packages tmLQCD~\cite{Jansen:2009xp,Abdel-Rehim:2013wba,Deuzeman:2013xaa,Kostrzewa:2022hsv},Lemon~\cite{Deuzeman:2011wz}, and QUDA~\cite{Clark:2009wm,Babich:2011np,Clark:2016rdz} have been used.

\bibliographystyle{JHEP}
\bibliography{refs} 

\clearpage

\appendix \label{appendix:A}

\section{Lattice data}
\begin{table}[ht]
 \caption{Pion and kaon masses in lattice units with statistical errors on the individual gauge ensembles. The superscript $i=1,2,3$ for $M_K^i$ labels results corresponding to the three values of $\mu^\mathrm{val,i}_s$ at each value of $\beta$, cf. Table~\ref{tab:beta_dependent_parameters}.}
 \centering
 \setlength{\tabcolsep}{0.5em}
 \begin{tabular}{lccccccc}
  \hline\hline
  ID           & $aM_\pi$ & $aM_K^{1}$ & $aM_K^{2}$ & $aM_K^{3}$ & $aM_{\eta_s}^{1}$ & $aM_{\eta_s}^{2}$ & $aM_{\eta_s}^{3}$\\
  \hline\hline
  cA211.12.48  & 0.08011(26) & 0.21732(45) & 0.24078(56) & 0.26210(68) & 0.32893(36) & 0.36115(33) & 0.39131(30) \\
  cA211.15.48  & 0.08901(05) & 0.21895(08) & 0.24243(10) & 0.26388(11) & 0.33028(21) & 0.36238(18) & 0.39245(16) \\
  cA211.30.32  & 0.12518(19) & 0.22775(16) & 0.25054(17) & 0.27150(19) & 0.33208(25) & 0.36406(23) & 0.39403(21) \\
  cA211.40.24  & 0.14508(29) & 0.23431(30) & 0.25665(31) & 0.27726(33) & 0.33350(36) & 0.36581(32) & 0.39603(30) \\
  cA211.53.24  & 0.16654(28) & 0.24140(31) & 0.26317(32) & 0.28335(34) & 0.33393(60) & 0.36624(54) & 0.39634(49) \\
  \hline
  cB211.072.64 & 0.05650(08) & 0.18050(24) & 0.20054(32) & 0.21883(40) & 0.26766(10) & 0.29563(09) & 0.32169(08) \\
  cB211.14.64  & 0.07834(11) & 0.18503(19) & 0.20488(24) & 0.22306(31) & 0.26847(15) & 0.29637(14) & 0.32236(13) \\
  cB211.25.48  & 0.10436(12) & 0.19124(15) & 0.21048(17) & 0.22817(19) & 0.26935(14) & 0.29726(12) & 0.32329(11) \\
  cB211.25.32  & 0.10536(28) & 0.19204(25) & 0.21131(26) & 0.22905(28) & 0.26853(37) & 0.29661(32) & 0.32275(29) \\
  cB211.25.24  & 0.10996(50) & 0.19407(73) & 0.21327(74) & 0.23097(76) & 0.27111(61) & 0.29892(54) & 0.32484(50) \\
  \hline
  cC211.06.80  & 0.04748(10) & 0.15318(27) & 0.17057(36) & 0.18589(46) & 0.22118(08) & 0.24652(07) & 0.26927(07) \\
  cC211.20.48  & 0.08588(23) & 0.16184(19) & 0.17880(21) & 0.19391(24) & 0.22239(22) & 0.24764(20) & 0.27034(19) \\
  \hline
  cD211.054.96 & 0.04056(07) & 0.14274(21) & 0.14471(22) & 0.14951(23) & 0.20248(06) & 0.20543(06) & 0.21267(06) \\
  \hline\hline
 \end{tabular}
 \label{tab:M_pi_M_K_and_M_eta_s}
\end{table}

\begin{table}[ht]
 \caption{$\eta$ and $\eta'$ masses in lattice units with statistical errors on the individual gauge ensembles. The superscript $i=1,2,3$ for $M_\eta^i$ and $M_{\eta'}^i$ labels results corresponding to the three values of $\mu^\mathrm{val,i}_s$ at each value of $\beta$, cf. Table~\ref{tab:beta_dependent_parameters}.}
 \centering
 \setlength{\tabcolsep}{0.5em}
 \begin{tabular}{lcccccc}
  \hline\hline
  ID           & $aM_\eta^{1}$ & $aM_\eta^{2}$ & $aM_\eta^{3}$ & $aM_{\eta'}^{1}$ & $aM_{\eta'}^{2}$ & $aM_{\eta'}^{3}$ \\
  \hline\hline
  cA211.12.48  & 0.2807(36) & 0.2988(53) & 0.3132(72) & 0.4382(140) & 0.4438(125) & 0.4533(109) \\
  cA211.15.48  & 0.2851(16) & 0.3048(26) & 0.3198(38) & 0.4456(38)  & 0.4541(34)  & 0.4659(30)  \\
  cA211.30.32  & 0.2910(14) & 0.3102(24) & 0.3253(36) & 0.4611(69)  & 0.4666(63)  & 0.4752(56)  \\
  cA211.40.24  & 0.2979(06) & 0.3177(09) & 0.3335(13) & 0.4658(78)  & 0.4702(72)  & 0.4780(66)  \\
  cA211.53.24  & 0.3009(08) & 0.3211(11) & 0.3373(15) & 0.4683(70)  & 0.4726(65)  & 0.4806(58)  \\
  \hline
  cB211.072.64 & 0.2293(28) & 0.2468(41) & 0.2603(57) & 0.3824(64)  & 0.3882(59)  & 0.3935(52)  \\
  cB211.14.64  & 0.2320(25) & 0.2509(37) & 0.2652(55) & 0.3904(67)  & 0.3947(62)  & 0.4019(57)  \\
  cB211.25.48  & 0.2347(14) & 0.2520(23) & 0.2644(33) & 0.3963(60)  & 0.3997(56)  & 0.4040(51)  \\
  cB211.25.32  & 0.2365(17) & 0.2545(29) & 0.2707(45) & 0.3896(64)  & 0.3938(59)  & 0.3997(53)  \\
  cB211.25.24  & 0.2402(18) & 0.2584(30) & 0.2739(42) & 0.3922(92)  & 0.3971(87)  & 0.4010(77)  \\
  \hline
  cC211.06.80  & 0.1898(24) & 0.2087(37) & 0.2174(53) & 0.3259(47)  & 0.3305(43)  & 0.3463(42)  \\
  cC211.20.48  & 0.1950(12) & 0.2121(19) & 0.2254(26) & 0.3439(94)  & 0.3501(86)  & 0.3515(78)  \\
  \hline
  cD211.054.96 & 0.1712(27) & 0.1719(31) & 0.1772(35) & 0.2800(61)  & 0.2797(62)  & 0.2813(60)  \\
  \hline\hline
 \end{tabular}
 \label{tab:M_eta_and_M_etap}
\end{table}

\begin{table}[ht]
 \caption{Values of the valence strange-quark masses $a\mu_s^{\mathrm{val},i}$ used in the computation of the mass of the $\Omega$ baryon and the corresponding measurements of $aM_\Omega^i$. Note that the $\mu_s^{\mathrm{val},i}$ values are different from the ones in Table~\ref{tab:beta_dependent_parameters} which have been used in the remaining analysis.}
 \centering
 \setlength{\tabcolsep}{0.5em}
 \begin{tabular}{lcccccc}
  \hline\hline
  ID           & $a\mu_s^{\mathrm{val},1}$ &  $a\mu_s^{\mathrm{val},2}$ & $a\mu_s^{\mathrm{val},3}$ & $aM_\Omega^{1}$ & $aM_\Omega^{2}$ & $aM_\Omega^{3}$ \\
  \hline\hline
  cA211.12.48  & 0.0182 & 0.0227 & 0.0273 & 0.7854(43) & 0.8220(34) & 0.8583(28) \\
  cA211.30.32  & 0.0182 & 0.0227 & 0.0273 & 0.7893(22) & 0.8257(18) & 0.8615(15) \\
  cA211.40.24  & 0.0182 & 0.0227 & 0.0273 & 0.8004(52) & 0.8374(39) & 0.8729(31) \\
  \hline                                                                        
  cB211.072.64 & 0.0170 & 0.0195 & 0.0220 & 0.6814(13) & 0.7013(12) & 0.7209(11) \\
  \hline                                                                        
  cC211.06.80  & 0.0150 & 0.0170 & 0.0190 & 0.5792(16) & 0.5945(15) & 0.6095(14) \\
  \hline                                                                        
  cD211.054.96 & 0.0130 & 0.0140 & 0.0150 & 0.4835(22) & 0.4940(19) & 0.5051(17) \\
  \hline\hline
 \end{tabular}
 \label{tab:M_Omega}
\end{table}

\begin{table}[ht]
 \caption{Pion and kaon decay constants in lattice units with statistical errors on the individual gauge ensembles. The superscript $i=1,2,3$ for $f_K^i$ labels results corresponding to the three values of $\mu^\mathrm{val,i}_s$ at each value of $\beta$, cf. Table~\ref{tab:beta_dependent_parameters}.}
 \centering
 \setlength{\tabcolsep}{0.5em}
 \begin{tabular}{lcccc}
  \hline\hline
  ID           & $af_\pi$ & $af_K^{1}$ & $af_K^{2}$ & $af_K^{3}$ \\
  \hline\hline
  cA211.12.48  & 0.06187(16) & 0.07293(36) & 0.07468(47) & 0.07617(59)  \\
  cA211.15.48  & 0.06306(05) & 0.07366(06) & 0.07562(06) & 0.07743(07)  \\
  cA211.30.32  & 0.06691(10) & 0.07597(08) & 0.07800(08) & 0.07983(08)  \\
  cA211.40.24  & 0.06840(16) & 0.07698(12) & 0.07904(12) & 0.08090(13)  \\
  cA211.53.24  & 0.07150(17) & 0.07883(15) & 0.08086(15) & 0.08271(15)  \\
  \hline
  cB211.072.64 & 0.05266(07) & 0.06167(24) & 0.06321(32) & 0.06453(43)  \\
  cB211.14.64  & 0.05490(09) & 0.06341(18) & 0.06516(25) & 0.06676(33)  \\
  cB211.25.48  & 0.05772(10) & 0.06481(12) & 0.06645(13) & 0.06795(15)  \\
  cB211.25.32  & 0.05656(15) & 0.06432(13) & 0.06602(14) & 0.06757(15)  \\
  cB211.25.24  & 0.05265(30) & 0.06232(22) & 0.06414(23) & 0.06578(24)  \\
  \hline
  cC211.06.80  & 0.04499(07) & 0.05251(25) & 0.05351(36) & 0.05436(48)  \\
  cC211.20.48  & 0.04891(12) & 0.05507(11) & 0.05650(11) & 0.05776(12)  \\
  \hline
  cD211.054.96 & 0.03776(04) & 0.04491(24) & 0.04502(25) & 0.04526(27)  \\
  \hline\hline
 \end{tabular}
 \label{tab:f_pi_and_f_K}
\end{table}

\begin{table}[ht]
 \caption{Mixing angle and its statistical error on the individual gauge ensembles. The superscript $i=1,2,3$ for $\phi^i$ labels results corresponding to the three values of $\mu^\mathrm{val,i}_s$ at each value of $\beta$, cf. Table~\ref{tab:beta_dependent_parameters}.}
 \centering
 \setlength{\tabcolsep}{0.5em}
 \begin{tabular}{lccc}
  \hline\hline
  ID           & $\phi^{1}/\deg$ & $\phi^{2}/\deg$ & $\phi^{3}/\deg$ \\
  \hline\hline
  cA211.12.48  & 42.7(2.0) & 37.5(2.3) & 32.1(2.4) \\
  cA211.15.48  & 43.5(0.8) & 38.7(1.0) & 33.7(1.0) \\
  cA211.30.32  & 44.5(0.7) & 39.7(0.8) & 34.6(0.9) \\
  cA211.40.24  & 43.6(0.4) & 38.7(0.4) & 33.8(0.5) \\
  cA211.53.24  & 45.2(0.5) & 40.4(0.6) & 35.7(0.6) \\
  \hline
  cB211.072.64 & 44.0(1.2) & 39.3(1.4) & 34.2(1.5) \\
  cB211.14.64  & 46.1(1.0) & 41.6(1.1) & 36.6(1.2) \\
  cB211.25.48  & 47.2(0.6) & 42.9(0.7) & 37.9(0.8) \\
  cB211.25.32  & 45.5(0.7) & 40.7(0.9) & 35.6(1.1) \\
  cB211.25.24  & 46.2(1.0) & 41.7(1.2) & 36.4(1.4) \\
  \hline
  cC211.06.80  & 46.2(1.1) & 41.5(1.3) & 36.1(1.4) \\
  cC211.20.48  & 45.3(0.7) & 40.5(0.9) & 35.6(1.0) \\
  \hline
  cD211.054.96 & 38.4(1.4) & 37.9(1.5) & 36.0(1.5) \\
  \hline\hline
 \end{tabular}
 \label{tab:phi}
\end{table}

\begin{table}[ht]
 \caption{Decay-constant parameters in lattice units with statistical errors on the individual gauge ensembles. The superscript $i=1,2,3$ for $f_{l,s}^i$ labels results corresponding to the three values of $\mu^\mathrm{val,i}_s$ at each value of $\beta$, cf. Table~\ref{tab:beta_dependent_parameters}.}
 \centering
 \setlength{\tabcolsep}{0.5em}
 \begin{tabular}{lcccccc}
  \hline\hline
  ID           & $af_l^{1}$ & $af_l^{2}$ & $af_l^{3}$ & $af_s^{1}$ & $af_s^{2}$ & $af_s^{3}$ \\
  \hline\hline
  cA211.12.48  & 0.0607(30) & 0.0628(30) & 0.0648(28) & 0.0804(18) & 0.0810(24) & 0.0792(31) \\
  cA211.15.48  & 0.0617(14) & 0.0636(14) & 0.0652(13) & 0.0827(07) & 0.0833(10) & 0.0821(13) \\
  cA211.30.32  & 0.0656(10) & 0.0680(11) & 0.0700(10) & 0.0851(06) & 0.0855(09) & 0.0840(12) \\
  cA211.40.24  & 0.0668(07) & 0.0693(07) & 0.0711(07) & 0.0851(05) & 0.0855(06) & 0.0843(07) \\
  cA211.53.24  & 0.0696(08) & 0.0722(09) & 0.0741(08) & 0.0866(07) & 0.0874(07) & 0.0868(08) \\
  \hline
  cB211.072.64 & 0.0523(15) & 0.0536(15) & 0.0547(14) & 0.0701(09) & 0.0708(13) & 0.0699(17) \\
  cB211.14.64  & 0.0523(11) & 0.0539(12) & 0.0555(11) & 0.0724(08) & 0.0735(10) & 0.0730(14) \\
  cB211.25.48  & 0.0545(06) & 0.0559(07) & 0.0573(07) & 0.0736(05) & 0.0750(07) & 0.0750(09) \\
  cB211.25.32  & 0.0553(07) & 0.0575(08) & 0.0594(08) & 0.0716(06) & 0.0719(08) & 0.0711(11) \\
  cB211.25.24  & 0.0496(11) & 0.0513(11) & 0.0528(10) & 0.0722(09) & 0.0734(12) & 0.0725(14) \\
  \hline
  cC211.06.80  & 0.0433(11) & 0.0451(11) & 0.0465(10) & 0.0610(07) & 0.0619(10) & 0.0613(13) \\
  cC211.20.48  & 0.0471(06) & 0.0490(07) & 0.0505(07) & 0.0610(05) & 0.0614(07) & 0.0605(08) \\
  \hline
  cD211.054.96 & 0.0419(09) & 0.0420(09) & 0.0425(09) & 0.0496(11) & 0.0496(12) & 0.0492(13) \\
  \hline\hline
 \end{tabular}
 \label{tab:f_l_and_f_s}
\end{table}

\end{document}